\documentclass[fleqn,12pt]{article}
\usepackage{amssymb,graphics,epsfig,rotating,cite,subfigure,sidecap,amsmath}
\def\beq   {\begin{equation}}
\def\eeq   {\end{equation}}
\def\beqd  {\begin{displaymath}}
\def\eeqd  {\end{displaymath}}
\def\beqaa {\begin{eqnarray}}
\def\eeqaa {\end{eqnarray}}

\def\ti  {\tilde}

\def\sq  {\ti q}
\def\sb  {\ti b}
\def\sg  {\ti g}

\def\sl  {\ti \ell}

\def\chm {\tilde\chi^-}

\def\t   {\theta}

\def\orp {\overrightarrow{p}}

\def\sz{\ifmmode{\tilde{\chi}^0} \else{$\tilde{\chi}^0$} \fi}
\def\sw{\ifmmode{\tilde{\chi}} \else{$\tilde{\chi}$} \fi}

\newcommand{\be}[1]{\begin{equation} \label{(#1)}}
\newcommand{\ee}{\end{equation}}
\newcommand{\baq}[1]{\begin{eqnarray} \label{(#1)}}
\newcommand{\eaq}{\end{eqnarray}}

\newcommand{\ba}{\begin{array}}
\newcommand{\ea}{\end{array}}

\newcommand{\neu}[1]{\tilde{\chi}^0_{#1}}
\newcommand{\cha}{\tilde{\chi}}
\newcommand{\st}{\tilde{t}}
\newcommand{\mneu}[1]{m_{\tilde{\chi}^0_{#1}}}
\newcommand{\mcha}[1]{m_{\tilde{\chi}^\pm_{#1}}}
\newcommand{\sbot}{\tilde{b}}

\begin{document}
\pagestyle{empty}

\begin{flushright}
DCPT-10-126 \\
DESY 10-094\\
IPPP-10-63 \\
\end{flushright}

\vfill

\begin{center}

{\Large {\bf
Momentum reconstruction at the LHC for probing CP-violation in the stop sector
}}

\vspace{10mm}

{\large
G.~Moortgat-Pick$^{a,c}$, K.~Rolbiecki$^{b,c}$, J.~Tattersall$^{b,d}$
}

\vspace{6mm}

\begin{center}
$^a${\it II. Institut fuer Theoret. Physik, University of Hamburg , Luruper Chaussee 149, D-22761 Hamburg, Germany}\\
$^b${\it IPPP, University of Durham, Durham DH1 3LE, UK}\\
$^c${\it DESY, Deutsches Elektronen-Synchrotron, Notkestr. 85, D-22607 Hamburg, Germany} \\
$^d${\it Bethe Centre for Theoretical Physics \& Physikalisches Institut, Universit\"{a}t Bonn,
D-53115 Bonn, Germany}
\end{center}

\end{center}

\vspace{1cm}

\begin{abstract}

We study the potential to observe CP-violating effects in supersymmetric $\tilde{t}_1$-cascade
decay chains at the LHC. Asymmetries composed of triple products of
momenta of the final state particles are sensitive to CP-violating
effects. Due to large boosts that dilute the asymmetries, these can be
difficult to observe. If all particle masses in a cascade decay are
known, it may be possible to reconstruct all momenta in the decay chains on an
event-by-event basis even when we have missing momentum due to a stable 
lightest supersymmetric particle. After the reconstruction, the non-diluted CP-violating signal can be recovered and gets significantly enhanced so that an observation may become feasible. A fully
hadronic study has been completed to define the areas of the
mSUGRA parameter space that may yield a 3-$\sigma$ observation with 500~fb$^{-1}$ at the LHC.

\end{abstract}

\vfill

\begin{flushleft}
e-mail: \texttt{\\ gudrid.moortgat-pick@desy.de\\ 
krzysztof.rolbiecki@durham.ac.uk \\ jamie.tattersall@durham.ac.uk}
\end{flushleft}

\newpage
\pagestyle{plain}
\tableofcontents
\newpage
\section{Introduction}\label{sec:introduction}

The Minimal Supersymmetric Standard Model (MSSM) is a particularly
compelling extension of the Standard Model, that may soon be explored
at the Large Hadron Collider (LHC).  It allows one to stabilize the
hierarchy between the electroweak (EW) scale and the Planck scale and
to naturally explain electroweak symmetry breaking (EWSB) by a
radiative mechanism. The naturalness of the scale of electroweak
symmetry breaking and the Higgs mass places a rough upper bound on the
superpartner masses of several TeV and the fits to the electroweak
precision data point to a rather light SUSY
spectrum~\cite{Buchmueller:2009fn}. If supersymmetry is discovered,
many studies will be required to determine the exact details of its
realisation. One of the interesting issues in this context is CP
violation. While the observed amount of CP violation in the $K$ and
$B$ sectors can be accommodated within the SM, another piece of
evidence, the baryon asymmetry of the universe, requires a new source
of CP violation~\cite{Cohen:1993nk,Gavela:1994dt,Rubakov:1996vz}.

The MSSM contains 105 free parameters
\cite{Dimopoulos:1995ju} and a large number of these may have non-zero
CP-violating phases, see e.g.~Ref.\cite{Ibrahim:2007fb}. Many of the
phases are unphysical in the sense that they can be rotated away by a
redefinition of the fields. The parameters normally chosen to be
complex and relevant to this study are the U(1) and SU(3) gaugino mass
parameters $M_1$ and $M_3$, the higgsino mass parameter $\mu$ and the
trilinear couplings of the third generation sfermions $A_f$
($f=b,t,\tau$). Hence we have,
\begin{equation}
M_1 = |M_1| e^{i \phi_1}\,, \qquad M_3 = |M_3| e^{i \phi_3}\,, \qquad
\mu = |\mu| e^{i \phi_\mu} \,, \qquad A_f = |A_f| e^{i \phi_{A_f}} \,.
\end{equation}
The two complex parameters that enter the $\st$ sector at tree level
are $A_t$ and $\mu$ and in the $\tilde{\chi}^0_i$ sector $\mu$ and $\phi_{1}$.
Certain combinations of the CP-violating phases
of these parameters are constrained by the experimental upper bounds
on various electric dipole moments (EDMs), see
e.g.~Ref.\cite{Ellis:2008zy}. Ignoring possible cancellations, the most
severely constrained phase is that of $\mu$ which contributes to the
EDMs at the one-loop level. In general for $\mathcal{O}(100)$~GeV
supersymmetric masses, $|\phi_{\mu}|$ has to be very small and we therefore set
$\phi_{\mu}=0$ throughout our study. The phase of $A_t$ has weaker
constraints as it only contributes to the EDMs at the two loop
level\cite{Kizukuri:1992nj,Ibrahim:1998je,Ibrahim:1999af,Brhlik:1998zn,Abel:2001vy,Arnowitt:2001pm,Li:2010ax}. Here
we study the complete range of $\phi_{A_t}$ in order to see the general
dependencies exhibited by our observables and the luminosity required
to observe this within the LHC environment. In principle, $\phi_{1}$ can also contribute to our observables but in the mSUGRA scenarios discussed in this paper, the dependence is weak due to the wino character of the $\tilde{\chi}^0_2$. We would like to
stress that in the chosen scenario experimental bounds from EDMs can
be evaded by arranging cancellations between various supersymmetric
contributions for any value of
$A_t$~\cite{Lee:2003nta,Lee:2007gn,Ellis:2008zy,Deppisch:2009nj}.

In general CP phases alter the couplings and masses of SUSY particles,
see Ref.\cite{Kraml:2007pr} for a recent review at the LHC. Therefore, in
principle we could detect CP-violating effects by studying mass
spectra, cross sections and branching ratios \cite{Barger:1999tn,Kneur:1999nx}.
However, to interprete these measurements accurately, we will require high
precision and will rely on many assumptions of the underlying SUSY breaking mechanism.
In addition, all of these
observables are CP-even and can be faked by a multitude of other
parameters.

In order to make the unambiguous observation of a complex parameter,
we need to use CP-odd observables. Examples of CP-odd observables
include rate asymmetries of cross sections and branching ratios. Another possibility, however, are observables
that are odd under T-transformations. Applying CPT-invariance, T-odd observables can be transferred under 
certain conditions into 
CP-odd variables, see Sec.~\ref{sec:Todd_Struc}. These kinds of observables
 can be defined using the triple product correlations of momenta
that are based on spin correlations of particles, see~Refs.\cite{Kittel:2009fg,Hesselbach:2007dq}
for a recent review. For the case of SUSY at the LHC, we can do this
using the final state particles of cascade decays.

The investigation of triple product correlations within SUSY at the LHC
has been looked at for various different processes. Stop cascade decays were
first studied in~Ref.\cite{Bartl:2004jr} and large CP-violating asymmetries
were found. The study shown in~Ref.\cite{Langacker:2007ur} was the first to
specifically examine stop decays in relation to the LHC and significant
dilution factors were noted when parton distribution functions were introduced. Explicit dilution
factors and initial estimates of the luminosity required at the LHC were
shown for three body decays in~Ref.\cite{Ellis:2008hq} and two body decays in~Ref.\cite{Deppisch:2009nj}.
In addition $\tilde{t}_2$ decays were investigated in~Ref.\cite{Kiers:2006aq}. $\sb$ decays have also been looked at in similar studies for two body cascade decays \cite{Bartl:2006hh,Deppisch:2010nc}. 

In~Ref.\cite{MoortgatPick:2009jy} we
looked at $\sq \sg$ production and decay and studied how to cope with statistical limitations and dilution factors
in searching for CP-phases in SUSY at the LHC. For the present paper, we extend the ideas described in detail
in~Ref.\cite{MoortgatPick:2009jy} of momentum reconstruction to $\st$
production and two-body decays. We further include hadronic,
combinatorial and background effects to study whether CP-violation
will be observable in the $\st$ sector at the LHC.

Regarding a precise measurement of $\phi_{A_t}$ at a future linear collider, we are not aware of any studies that measure CP-violation in the stop sector directly. However, studies have been completed to measure the absolute value of $A_t$ and are expected to be accurate to within 10\% \cite{Boos:2003vf,Weiglein:2004hn}. Also, there is the potential to study the CP properties of other 3rd generation trilinear couplings, namely those of the $\tilde{\tau}$ sector, $\phi_{A_{\tau}}$ \cite{Dreiner:2010ib,Dreiner:2010wj}. In addition, these studies may also be applicable to the LHC if tau polarisation can be probed in the final state \cite{Nattermann:2009gh}.

For our study, we consider the LHC production process,
\begin{equation}
  pp\to \st_1 \st_1^*. 
\end{equation}
Our signal CP-odd observable is generated in the following two body decays,
\begin{equation}
  \st_1 \to \neu{2} t, \qquad \neu{2} \to \sl \ell_N, \qquad \sl \to \neu{1} \ell_F,  \qquad t \to b + W. \label{eq:SimpDec}
\end{equation}
where $\ell_N$ and $\ell_F$ denote the near and far leptons respectively. The CP-odd observables are built from triple products of final state momentum or
reconstructable particles, e.g. $\vec{p}_{\ell_{N}} \cdot (\vec{p}_{t} \times \vec{p}_{W})$. 

Triple products constructed in this way are not Lorentz invariant but
instead depend on the intrinsic boost of the produced particle in the
laboratory frame. The observed asymmetry is maximal when the decay is
at rest in the laboratory frame and any boost dilutes the
observable. Consequently, we decided to use the idea of momentum
reconstruction to find the momentum of the invisible
$\neu{1}$. We are able to perform momentum reconstruction for the
decay chain shown in Eq.~(\ref{eq:SimpDec}) as we have four on-shell
mass conditions which we can solve for the four unknowns of the $\neu{1}$
momentum on an event-by-event basis. Once the $\neu{1}$ momentum is known, we can find the rest
frame of any particle involved in the decay chain and thus measure the
maximum CP asymmetry.

An important note to make is that the sign of the asymmetry generated
by the triple product flips if we consider the decay of the charge
conjugate $\st^*_1$. Therefore, in addition to measuring the triple
product we must also determine the charge of the decaying
$\st_1$. Unfortunately we cannot use a leptonically decaying $W$ in
this study as we must fully measure the $t$ momentum to perform
momentum reconstruction. Hence, we rely on the opposite $\st_1$ decay
to a single charged lepton final state to tag the charge of both
produced stops e.g. $\st_1^* \to \neu{1} \bar{t},\, \bar{t} \to \bar{b} \ell^-
\bar{\nu}_{\ell}$. As an aside, charge identification of the process is also required to
rule out T$_N$-odd observables that can in principle be generated by final
state interactions at the one-loop level \cite{Atwood:2000tu}, see Sec.~\ref{sec:Todd_Struc} for more details. We
compare the signal process with the charge-conjugated decay and if a
non-zero asymmetry is observed in the combination, it must correspond
to a violation of CP symmetry.

Apart from backgrounds due to hard interactions, measuring asymmetries in a hadronic environment is challenging due to the high QCD activity and underlying event that can be hard to disentangle from the signal process. However, D0 at the Tevatron has succeeded in making such a measurement with the like-sign dimuon charge asymmetry \cite{Abazov:2010hv}. The asymmetry is interpreted to originate from the mixing of neutral B mesons and differs by 3.2 standard deviations from the standard model prediction. Therefore, the measurement is a significant hint of a new source of CP-violation. Although this particular measurement is unlikely to be possible at the LHC due to the $pp$ initial state, it does show that if the correct observables are chosen, asymmetry observations are possible at hadron colliders. In addition, CDF at the Tevatron has also recently made the observation of an asymmetry in the pair production of top quarks that also hints at new physics \cite{Aaltonen:2011kc}.

We begin in Sec.~\ref{sec:formalism} by describing the process and underlying structure 
to derive the various triple products that can be
formed. In Sec.~\ref{sec:MomRec} we discuss the momentum
reconstruction method and its application to the process
studied. Sec.~\ref{sec:PartResults} gives the analytical results of
the asymmetries at parton level. Hadron level results are described in
Sec.~\ref{sec:HadResults} where we also discuss the effects of
standard model and SUSY backgrounds. We also find that the method of momentum reconstruction significantly improves the signal to background ratio.


\section{Formalism}
\label{sec:formalism}

\subsection{The process studied and the amplitude squared }
\label{sec:process}

\begin{figure}[ht!]
\begin{picture}(10,8)
  \put(-4,-17){\epsfig{file=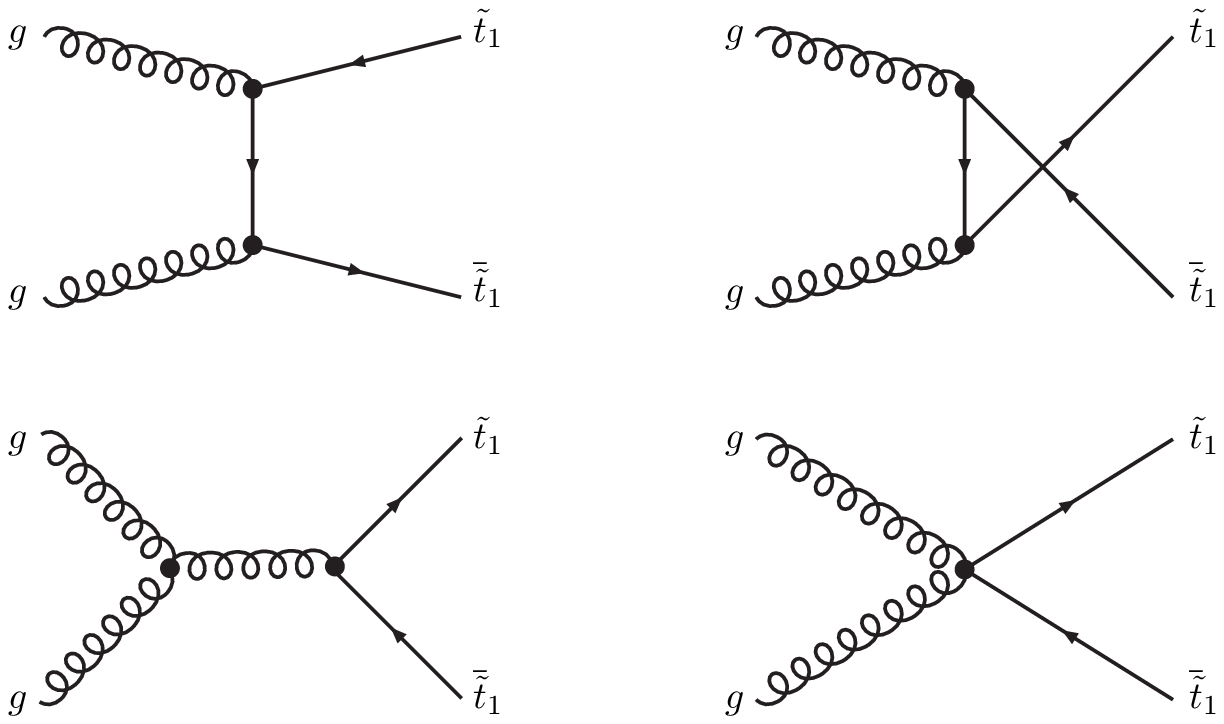,scale=1}}
\end{picture}
\caption{\label{Fig:FeynProd}Feynman diagrams for the production process
  $gg\to\st_1\overline{\st}_1$.}
\end{figure}

At the LHC, the light stop ($\st_1$) particles can be produced via pair production,
\begin{equation}
	pp\to\st_1\st^*_1\,.
\end{equation}
which at the LHC will be dominated by the gluon fusion channels, Fig.~\ref{Fig:FeynProd}.
In our study the CP-violating observables are produced in the following decay,
\begin{eqnarray}
\tilde{t}_1 \to \tilde{\chi}^0_2 t\,.
\label{st-decay} 
\end{eqnarray}
We require the $\tilde{\chi}^0_2$ to decay via two, 2-body leptonic channels,
\begin{eqnarray}
\tilde{\chi}^0_2 \to \tilde{\ell}^{\pm}_R \ell^{\mp}_{N} \to \tilde{\chi}^0_1 \ell^{\mp}_{N} \ell^{\pm}_{F}\,,
\end{eqnarray}
where N and F denote the near and far leptons respectively.
In addition, we only consider events where the $t$ is fully reconstructable and hence decays hadronically,
\begin{eqnarray}
t \to Wb \to q_u\bar{q}_db\,.
\end{eqnarray}

Using the formalism of~Refs.\cite{MoortgatPick:1999di,Haber:1994pe}, the
squared amplitude $|T|^2$ of the full process can be factorised into
the processes of production $gg\to \tilde{t}_1\tilde{t}_1^*$ and the
subsequent decays $\tilde{t}_1 \to t \tilde{\chi}^0_2$,
$\tilde{\chi}^0_2\to\tilde{\ell}\ell_N$,
$\tilde{\ell}\to\neu{1}\ell_F$ and $t\to W b$. We apply the
narrow-width approximation but include the full spin correlations for the production and
the decay of the intermediate particles, $\tilde{t}_1$,
$\tilde{\chi}^0_2$, $\tilde{\ell}$ and $t$. The use of the narrow-width approximation
is  appropriate since
the widths of the respective particles are much smaller than the masses
in all cases.
The squared amplitude can then be expressed 
in the
form,
\begin{eqnarray}
|T|^2 &=& 4 |\Delta(\tilde{t}_1)|^2 |\Delta(\tilde{\chi}^0_2)|^2 |\Delta(\tilde{\ell})|^2 |\Delta(t)|^2
 P(\tilde{t}_1\tilde{t}_1^*) \Big\{ P(\tilde{\chi}^0_2 t)
 D(\tilde{\chi}^0_2) D(\tilde{\ell}) D(t) \nonumber\\
      &&       +
            \sum^3_{a=1}\Sigma^a_P(\tilde{\chi}^0_2)
                    \Sigma^a_D(\tilde{\chi}^0_2)D(\tilde{\ell})D(t)
             +
            \sum^3_{b=1}\Sigma^b_P(t)
                    \Sigma^b_D(t)D(\tilde{\chi}^0_2)D(\tilde{\ell}) \nonumber\\
      &&       +
            \sum^3_{a,b=1}\Sigma^{ab}_P(\tilde{\chi}^0_2t)
                    \Sigma^a_D(\tilde{\chi}^0_2) \Sigma^b_D(t)D(\tilde{\ell})
\Big\},
\label{Tsquared}
\end{eqnarray}
where $a,b=1,2,3$ refers to the polarisation states of the neutralino
$\tilde{\chi}^0_i$ and top quark $t$. In addition,
\begin{itemize}
\item $\Delta(\tilde{t}_1)$, $\Delta(\tilde{\chi}^0_2)$, $\Delta(\tilde{\ell})$ and $\Delta(t)$ are the
  pseudo-propagators of the intermediate particles which lead to the factors
  $E_{\tilde{t}_1}/m_{\tilde{t}_1}\Gamma_{\tilde{t}_1}$,
  $E_{\tilde{\chi}^0_2}/m_{\tilde{\chi}^0_2}\Gamma_{\tilde{\chi}^0_2}$, $E_{\tilde{\ell}_R}/m_{\tilde{\ell}_R}\Gamma_{\tilde{\ell}_R}$ and $E_t/m_t\Gamma_{t}$ in the narrow-width 
approximation.
\item $P(\tilde{t}_1\tilde{t}_1)$, $P(t\tilde{\chi}^0_2)$, $D(\tilde{\chi}^0_{2})$, $D(\tilde{\ell})$
and $D(t)$ (Appendix~\ref{sect:AmpSq}) are the terms in the
  production and decay that are independent of the spin of the decaying 
  neutralino and top,  whereas,
\item $\Sigma^a_P(\tilde{\chi}^0_{i})$, $\Sigma^b_P(t)$,
  $\Sigma^{ab}_P(\tilde{\chi}^0_2t)$ and $\Sigma^a_D(\tilde{\chi}^0_{2})$, $\Sigma^b_D(t)$ (Appendix~\ref{sect:AmpSq})
  are the spin-dependent terms giving the correlations 
between production and decay of the $\tilde{\chi}^0_2$ and $t$. We follow the formalism and
conventions described in~Ref.\cite{Haber:1994pe}.
\item
	It must be noted that the slepton $\tilde{\ell}$ produces no spin correlation term in the amplitude since it is a scalar.
\end{itemize}

Explicit expressions are given in Appendix~\ref{sect:AmpSq}.

\subsection{Structure of the T-odd asymmetry}
\label{sec:Todd_Struc}

As shown in the CPT-theorem \cite{Schwinger:1951xk,Schwinger:1953tb}, relativistic quantum
field theories with usual spin-statistics relations have to be
invariant under a CPT-transformation. This invariance guarantees that
the masses and also the total widths of particles and antiparticles are
the same. Since a true T-transformation is anti-unitary, which exchanges the
initial and the final states, it is useful to study
'naive' T$_{N}$-transformations for collider-based experiments. The
definition of T$_{N}$-transformations is to apply T-transformations to
the initial and final states but without interchanging them.  The
unitarity of the S--matrix leads in the absence of re-scattering effects
(i.e. in leading order in perturbation theory, no final state interactions (FSI) and no width
effects) to a conservation of the scattering amplitude under a
CPT$_{N}$-transformation ~\cite{Atwood:2000tu}.

It is therefore useful to categorise CP--violating observables into
T$_{N}$--odd and T$_{N}$--even observables. CPT$_N$ invariance implies that a T$_{N}$-odd observable is also CP-odd in the absence of re-scattering effects. However, in case re-scattering effects contribute, i.e.\
CPT$_{N}\neq$ CPT-invariance, a T$_N$-odd signal may be caused by
such re-scattering effects and does not necessarily imply CP-violation.

For all our observables we require that we know the charge of the decaying $\st_1$ and can therefore distinguish the particle and anti-particle. Hence we can combine the process with the charge-conjugated decay to make an unambiguous observation of CP-violation via T$_N$-odd observables. 

In general, it is therefore important to classify all terms of the
corresponding amplitude squared, eq.(\ref{Tsquared}), with respect to
their T$_N$--odd or T$_N$--even character. Only the products that
contain a T$_N$-odd contribution will lead to CP-odd violating
observables:
\begin{itemize} 
\item The spin--independent terms introduced in the previous section, 
 $P(\tilde{t}_1\tilde{t}_1)$, $P(t\tilde{\chi}^0_2)$, $D(\tilde{\chi}^0_{2})$, 
$D(\tilde{\ell})$, $D(t)$ do not cause any T$_N$-odd terms.
\item The spin-dependent terms, 
$\Sigma^a_P(\tilde{\chi}^0_{i})$, $\Sigma^b_P(t)$,
  $\Sigma^{ab}_P(\tilde{\chi}^0_2t)$, 
$\Sigma^a_D(\tilde{\chi}^0_{2})$, $\Sigma^b_D(t)$,
however, often can be divided up into T$_N$-even and T$_N$-odd terms, depending on the 
processes \newpage studied. In our case, a sequence of 2-body decays, we can only split
$\Sigma^{ab}_P(\tilde{\chi}^0_2t)=\Sigma^{ab}_{P,even}(\tilde{\chi}^0_2t)
+\Sigma^{ab}_{P,odd}(\tilde{\chi}^0_2t)$ and other spin-dependent terms only lead to 
$T_N$-even terms\footnote{This is different if 3-body decays are studied, see~Ref.
\cite{Ellis:2008hq}. In that case spin-dependent terms from both the production  
$\Sigma^{ab}_P(\tilde{\chi}^0_2t)$ as well as from the 3-body decay 
$\Sigma^a_D(\tilde{\chi}^0_{2})$ lead to CP-odd contributions.}.

\item
Therefore, the T$_N$-odd term in the amplitude is, $\sum^3_{a,b=1}\Sigma^{ab}_{P,odd}(\tilde{\chi}^0_2t)
                    \Sigma^a_D(\tilde{\chi}^0_2) \Sigma^b_D(t)D(\tilde{\ell})$. 
 \end{itemize}

When we contract the spin indices of the $t$ and $\neu{2}$ and evaluate the T$_N$-odd contribution, we find that the following covariant product appears in the amplitude,
\begin{eqnarray}
\Sigma^{ab}_{P,odd}(\tilde{\chi}^0_2 t)
                    \Sigma^{a}_D(\tilde{\chi}^0_2)\Sigma^{b}_D(t) & \sim &
 i\epsilon_{\mu\nu\rho\sigma}s^{a,\mu}(\tilde{\chi}^0_2)p^{\nu}_{\tilde{\chi}^0_2}
s^{b,\rho}(t)p^{\sigma}_t \times (p_{\ell_N} s^{a})(p_{[b,W]}s^b), \\
& \sim & i\epsilon_{\mu\nu\rho\sigma} p^{\nu}_{\tilde{\chi}^0_2} p^{\mu}_{\ell_N} p^{\rho}_{W} p^{\sigma}_t\,,
\label{eq_term1}
\end{eqnarray}
where $\Sigma^{ab}_{P,odd}$, $\Sigma^{a}_D(\tilde{\chi}^0_2)$ and $\Sigma^{b}_D(t)$ are given by Eq.~(\ref{eq_prod-o}), Eq.~(\ref{eq_neutdecay}) and Eq.~(\ref{eq_stdecay}), respectively.

The above equation is multiplied by the imaginary part of the coupling, Eq.~(\ref{eq:ImCoup}) that contains terms from both the $\st$, Eq.~(\ref{eq:StopMix}), and $\tilde{\chi}^0$, Eq.~(\ref{eq:mass_matrix}), mixing matrices. Hence, any complex phases contained in those mixing matrices will yield CP-violating effects that can be seen in an observable that exploits the covariant product.
We can now expand the Lorentz invariant covariant product in terms of the explicit energy and momentum components,
\begin{eqnarray} 
  \epsilon_{\mu\nu\rho\sigma} p^{\nu}_{\tilde{\chi}^0_2} p^{\mu}_{\ell_N} p^{\rho}_{W} p^{\sigma}_t = & E_{\neu{2}}\;\overrightarrow{p_{\ell_N}}\cdot(\overrightarrow{p_{W}}\times\overrightarrow{p_{t}})+E_{W}\;\overrightarrow{p_{t}}\cdot(\overrightarrow{p_{\neu{2}}}\times\overrightarrow{p_{\ell_N}}) \label{eq:EpsExpanLab} \\ & -E_{\ell_N}\;\overrightarrow{p_{W}}\cdot(\overrightarrow{p_{t}}\times\overrightarrow{p_{\neu{2}}})-E_{t}\;\overrightarrow{p_{\neu{2}}}\cdot(\overrightarrow{p_{\ell_N}}\times\overrightarrow{p_{W}})\,. \nonumber 
\end{eqnarray}
The first term in Eq.~(\ref{eq:EpsExpanLab}) shows the CP sensitive triple product that can be measured from final state momenta. However, this triple product is not Lorentz invariant and consequently can vary in both magnitude and sign in different reference frames. If we are in the rest frame of the $\neu{2}$ though,
\begin{equation} 
  \epsilon_{\mu\nu\rho\sigma}p_{\neu{2}}^\mu p_{\ell_N}^\nu p_{W}^\rho p_{t}^\sigma \longrightarrow  m_{\neu{2}}\;\overrightarrow{p_{\ell_N}}\cdot(\overrightarrow{p_{W}}\times\overrightarrow{p_{t}})\,,
 \label{eq:EpsExpanRest}
\end{equation}
the resulting asymmetry, Eq.~(\ref{eq:Asy}), is uniquely defined since all other terms of the covariant product vanish as $\vec{p}_{\neu{2}}\to0$.

Hence we see that triple products of momenta, can be used as T$_N$-odd observables. In this paper we find that the triple products most useful to study are,
\begin{eqnarray}
  \mathcal{T}_{\ell_N} & = & \vec{p}_{\ell_{N}} \cdot (\vec{p}_{W} \times \vec{p}_{t})\,,  \label{eq:tpt} \\
 \mathcal{T}_{\ell\ell} & = & 
\vec{p}_b \cdot (\vec{p}_{\ell^+} \times \vec{p}_{\ell^-})\,.  \label{eq:tpb}
\end{eqnarray}
where ${\ell^+}$ and ${\ell^-}$ are the two leptons produced in the $\neu{2}$ cascade decay. For the triple product, Eq.~(\ref{eq:tpb}), the identification of near and far leptons is not required as is explained at the end of this section.

The T-odd asymmetry is then defined as,
\begin{eqnarray}
\label{eq:Asy}
\mathcal{A}_{T} = 
\frac{N_{\mathcal{T}_+}-N_{\mathcal{T}_-}}{N_{\mathcal{T}_+}+N_{\mathcal{T}_-}} &=&
\frac{\int\mathrm{sign}\{ \mathcal{T}_{f}\}
  |T|^2d\,\mbox{lips}}{{\int}|T|^2d\,\mbox{lips}}\,, 
 \end{eqnarray}
where $f=\ell_N$ or $\ell\ell$, $d\,\mbox{lips}$ denotes Lorentz invariant phase space and $N_{\mathcal{T}_+}$ ($N_{\mathcal{T}_-}$) are the numbers of events for which
$\mathcal{T}$ is positive (negative). The denominator in Eq.~(\ref{eq:Asy}), ${\int}|T|^2d\,\mbox{lips}$, is equal to the total
cross section.

We then define,
\begin{equation}
    \mathcal{A}_{\ell_N} = \mathcal{A}_{T}(\mathcal{T}_{\ell_N}),\qquad \qquad 
\mathcal{A}_{\ell\ell} = \mathcal{A}_{T}(\mathcal{T}_{\ell\ell}),
\end{equation}
where $\mathcal{A}_{\ell_N}$ is the asymmetry from the triple product 
$\mathcal{T}_{\ell_N}$ and 
$\mathcal{A}_{\ell\ell}$ is the asymmetry from the triple product $\mathcal{T}_{\ell\ell}$.

As stated above, whilst the covariant product is Lorentz invariant, the triple products are not. However, we can see that for the triple product in Eq.(\ref{eq:tpt}), the rest frame of the $\neu{2}$ and the $\st_1$ are equivalent since $(p_{\st}=p_{\neu{2}}+p_t)$,
\begin{equation} 
  \epsilon_{\mu\nu\rho\sigma}p_{\neu{2}}^\mu p_{\ell_N}^\nu p_{W}^\rho p_{t}^\sigma = \epsilon_{\mu\nu\rho\sigma}p_{\st_1}^\mu p_{\ell_N}^\nu p_{W}^\rho p_{t}^\sigma
 \label{eq:StNeuEquiv}\,.
\end{equation}
For the triple product $\mathcal{T}_{\ell\ell}$, Eq.~(\ref{eq:tpb}), the covariant product can be re-expressed in the following form (exploiting momentum conservation, $p_{\neu{2}}=p_{\sl}+p_{\ell_N}$, $p_{\sl}=p_{\ell_F}+p_{\neu{1}}$, $p_W=p_t+p_b$),
\begin{equation} 
  \epsilon_{\mu\nu\rho\sigma}p_{\neu{2}}^\mu p_{\ell_N}^\nu p_{W}^\rho p_{t}^\sigma = \epsilon_{\mu\nu\rho\sigma}(p_{\ell_F}+p_{\neu{1}})^\mu p_{\ell_N}^\nu p_{W}^\rho p_{b}^\sigma\,.
 \label{eq:EpsExpantpb}
\end{equation}
We now see that we have effectively two covariant products, one which contains the momentum of the $\neu{1}$. In general, triple products containing the momentum of the far lepton will be lower as the far lepton is not directly correlated with the spin of $\neu{2}$. Nevertheless, we can exploit and maximise the triple products originating from Eq.~(\ref{eq:StNeuEquiv}) and Eq.~(\ref{eq:EpsExpantpb}), if we know the momentum of the unstable particles in the decay chain. This can be provided by the momentum reconstruction procedure described in the following section.

Changing the decaying $\st_1$ to a $\st_1^*$ or changing the charge of the near lepton $\ell_N$ reverses the sign of the covariant product. Consequently we have to know the charge of both the $\st_1$ and the $\ell_N$, otherwise any asymmetry will cancel. The charge of the $\st_1$ can be found by demanding that the opposite cascade produces a single lepton and thus a tri-lepton final state. We distinguish the near and far leptons using the momentum reconstruction technique, Sec.~\ref{sec:MomRec}. However if for some reason the leptons cannot be identified, we can still use the triple product $\mathcal{T}_{\ell\ell}$, Eq.~(\ref{eq:tpb}). No lepton distinction is required as exchanging the near and far leptons has an extra sign change that cancels the change produced by the charge exchange.

\section{Momentum reconstruction}
\label{sec:MomRec}

\subsection{Dilution effects}

\begin{figure}[t!]
\begin{picture}(16,7.5)
\ \put(-2.5,-13){\epsfig{file=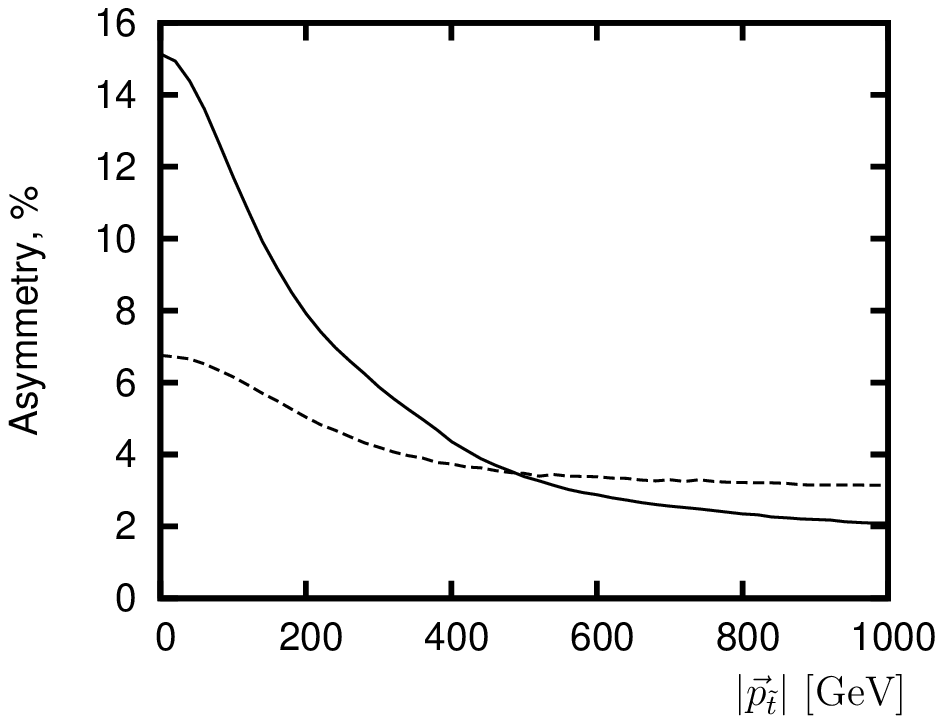,scale=1}}
\end{picture}
\caption{\label{fig:MomPlot} The asymmetry $\mathcal{A}_{\mathrm{T}}$, Eq.~(\ref{eq:Asy}), as a function of the stop momentum, $|\vec{p}_{\tilde{t}}|$, in the laboratory frame. The solid line is the asymmetry for the triple product $\mathcal{T}_{\ell_N}$, Eq.~(\ref{eq:tpt}), and the dotted line is for the triple product $\mathcal{T}_{\ell\ell}$, Eq.~(\ref{eq:tpb}). The respective masses are given in Tab.~\ref{tab:GaugeMasses}, Tab.~\ref{tab:SquarkMasses} and Tab.~\ref{tab:SelMasses}.}
\vspace{0.cm}
\end{figure}

The triple product that is constructed from momenta in the laboratory frame suffers from dilution factors ($\sim4$) at the LHC. This is due to the lab frame being boosted with respect to the rest frame of the $\tilde{\chi}^0_2$ or $\st_1$, see Eq.~(\ref{eq:StNeuEquiv}), for a more detailed discussion see~Ref.\cite{Ellis:2008hq}. It results in a considerable reduction in the maximum asymmetry observable when we introduce the parton density functions (PDFs) which causes an undetermined boost to the system. Fig.~\ref{fig:MomPlot} shows how the asymmetry is diluted in the laboratory frame when we produce the $\st_1$ with varying initial momenta. If we were able to reconstruct the momentum of the $\st_1$, we could perform a Lorentz transformation of all the momenta in the triple product into the $\st_1$ rest frame and potentially recover the full asymmetry.

\subsection{Reconstruction procedure}
\label{sec:RecProd}

We are able to reconstruct the $\tilde{\chi}^0_1$ four momentum by reconstructing the following two body decay chain in full,
\begin{equation}\label{eq:stopdecay}
 \tilde{t} \to t + \tilde{\chi}^0_2 \to t + \tilde{\ell}^{\pm} + \ell^{\mp}_N \to t + \tilde{\chi}^0_1 + \ell^{\mp}_N + \ell^{\pm}_F.
\end{equation}

\begin{SCfigure}[50]
  \begin{picture}(7,8.15)
 \put(-2.5,-13.5){\epsfig{file=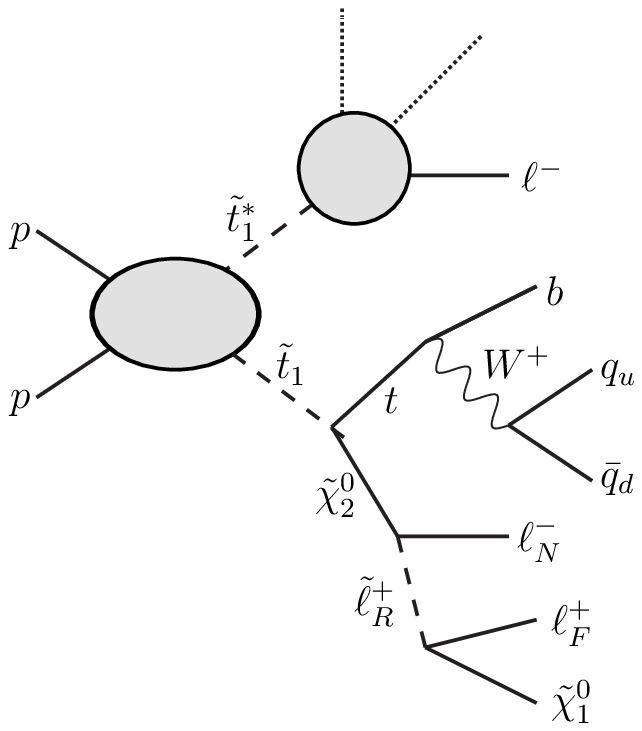,scale=1}}
 \end{picture}
   \hspace{0.8cm}   
   \caption{\label{fig:FullDiagram} The process studied for momentum reconstruction.}
\end{SCfigure}

Assuming that all the masses in the decay chain are known, the kinematics can be fully reconstructed using the set of invariant mass conditions,
\begin{eqnarray}
	 m^2_{\tilde{\chi}^0_1} & = & (p_{\tilde{\chi}^0_{1}})^2 \label{eq:chi1mass},\\
         m^2_{\tilde{\ell}^{\pm}} & = & (p_{\tilde{\chi}^0_{1}}+p_{\ell^{\pm}_F})^2, \label{eq:selmass} \\ 
	m^2_{\tilde{\chi}^0_2} & = & (p_{\sl^{\pm}}+p_{\ell^{\mp}_N})^2 = (p_{\tilde{\chi}^0_1}+p_{\ell^{\pm}_F}+p_{\ell^{\mp}_N})^2, \label{eq:chi2mass}\\
	m^2_{\tilde{t}_1} & = & (p_{\neu{2}}+p_{t})^2 = (p_{\tilde{\chi}^0_1}+p_{\ell^{\pm}_F}+p_{\ell^{\mp}_N}+p_t)^2 \label{eq:stopmass},
\end{eqnarray}
where $p$ denote the four momenta of the respective particles.

We see that with the four equations we have enough information to solve the system and find each component of the $\tilde{\chi}^0_1$ four momentum. A solution to the above set of equations is presented in~Ref.\cite{Kawagoe:2004rz} and we outline the procedure here. We first expand the $\neu{1}$ momentum in terms of the final state momentum of the $\ell^{\mp}_F$, $\ell^{\pm}_N$ and $t$,
\begin{equation}
	 \overrightarrow{p}_{\tilde{\chi}^0_{1}} = a\overrightarrow{p}_{\ell^{\pm}_F} + b\overrightarrow{p}_{\ell^{\mp}_N} + c\overrightarrow{p}_{t}\:. \label{eq:abc}
\end{equation}
In order to derive a system of 3 linear equations for the unknowns $a-c$, we calculate $\vec{p}_{\neu{1}}\cdot\vec{p}_{\ell_F}$, $\vec{p}_{\neu{1}}\cdot\vec{p}_{\ell_N}$ and $\vec{p}_{\neu{1}}\cdot\vec{p}_{t}$. Inserting Eq.~(\ref{eq:abc}) and exploiting Eqs.~(\ref{eq:selmass}-\ref{eq:stopmass}) we form the system of equations,
%
%
\begin{equation}
  \label{eq:MomMatEq}
  \mathcal{M}\left(
    \begin{array}{c}
      a \\ b \\ c 
    \end{array} \right)= \left(
    \begin{array}{c}
       \frac{1}{2}(m^2_{\tilde{\chi}^0_1}-m^2_{\tilde{\ell}}) + E_{\tilde{\chi}^0_{1}}E_{\ell_F}    \\
       \frac{1}{2}(m^2_{\tilde{\ell}}-m^2_{\neu{2}}) + p_{\ell_F}\cdot p_{\ell_N} + E_{\tilde{\chi}^0_{1}}E_{\ell_N}\\
       \frac{1}{2}(m^2_{\neu{2}}+m^2_{t}-m^2_{\tilde{t}_1}) + p_{\ell_F}\cdot p_{t} + p_{\ell_N}\cdot p_{t} +E_{\tilde{\chi}^0_{1}}E_{t} \\
    \end{array} \right)\:,
\end{equation}
where,
\begin{eqnarray}
 \mathcal{M} = \left(
    \begin{array}{ccc}
       \orp_{\ell_F}\cdot\orp_{\ell_F}\;\; & \orp_{\ell_F}\cdot\orp_{\ell_N}\;\; & \orp_{\ell_F}\cdot\orp_{t} \\
       \orp_{\ell_N}\cdot\orp_{\ell_F}\;\; & \orp_{\ell_N}\cdot\orp_{\ell_N}\;\; & \orp_{\ell_N}\cdot\orp_{t}  \\
       \orp_{t}\cdot\orp_{\ell_F}\;\; & \orp_{t}\cdot\orp_{\ell_N}\;\; & \orp_{t}\cdot\orp_{t} 
    \end{array} \right)\nonumber. \\ 
\end{eqnarray}
We invert the matrix $\mathcal{M}$ to find solutions for $a$, $b$ and $c$ in terms of constants and $E_{\tilde{\chi}^0_{1}}$. The on shell mass condition for the $\neu{1}$, Eq.~(\ref{eq:chi1mass}), can then be expressed as,
\begin{equation}
	E^2_{\neu{1}} = (a,b,c)\mathcal{M}\left(
    \begin{array}{c}
      a \\ b \\ c 
    \end{array} \right) + m^2_{\neu{1}}.
\end{equation}
We solve the above quadratic equation, to find two solutions for $E_{\neu{1}}$. These solutions are then substituted back into Eq.~(\ref{eq:abc}) to find all components of the $\st_1$ momentum on an event-by-event basis.

\subsection{Challenges from multiple solutions}
\label{sec:MomRecPrac}

We encounter a complication in the reconstruction as
Eq.~(\ref{eq:chi1mass}) is quadratic in
$p_{\tilde{\chi}^0_{1}}$. Consequently we have two solutions for
$p_{\tilde{\chi}^0_{1}}$, for each reconstructed event but we have no
extra information in the single decay chain to determine which
solution is physically correct. As we cannot distinguish which of
these solutions corresponds to the physically correct configuration,
we need to analyse both. Therefore, we calculate the $\st_1$ momentum
for both configurations and boost all final state particles in the
triple product into the reconstructed $\st_1$ rest frame. If the sign of
both triple products are the same then the event is recorded but if
the sign of the triple products are different, we discard the event
since we cannot know which of the reconstructed solutions is
correct. The method has the disadvantage that we lose events and
therefore statistical significance. However, we find that the
asymmetry can actually rise ($\approx 1.5\%$) as most of the events
removed have small triple products and events with a small triple
product lead to smaller asymmetries.

The procedure is essentially a cut designed for the triple product correlation observables. Events with an ambiguous triple product sign will significantly dilute the asymmetry and reduce the statistical significance of any CP-violating observation. Therefore, they must be removed from the sample. The disadvantage of the cut is that it makes an actual measurement of the CP-violating phase more involved. A comparison would have to be performed between a Monte Carlo simulation and the real data for a measurement to take place and may induce new errors. However, we believe that an actual determination of the phases at the LHC will be challenging and the method presented is more designed to establish the presence of CP-violation in SUSY.

When performing the momentum reconstruction at the LHC we have
additional problems from multiple solutions that come from
combinatorial effects in the event. First, to complete the
reconstruction we need to correctly identify the near and far lepton
in the decay chain Eq.~(\ref{eq:stopdecay}), if we wish to compute the triple product $\mathcal{T}_{\ell_N}$, Eq.~(\ref{eq:tpt}), although this information is not required for the triple product $\mathcal{T}_{\ell\ell}$, Eq.~(\ref{eq:tpb}).
We find that in $\approx 20\%$ of events the wrong assignment of near and
far leptons satisfies the kinematic equations
Eq.~(\ref{eq:chi1mass})-(\ref{eq:stopmass}) and produces two extra
solutions for the momentum of the $\tilde{\chi}^0_1$ in addition to
the solutions found from the correct configuration. In addition, we
always require a third lepton in the event coming from the opposite
decay chain to correctly identify the stop charge. For example the
lepton produced in the decay chain $\st_1^* \to \chm \bar{b}$, $\chm \to
\ell^- + X$, where X are other neutral decay products . If this lepton
is of the same flavour as those in the triple product decay chain
there is a small chance that it can also reconstruct the
$\tilde{\chi}^0_2$. All of these combinatorial issues are removed by
again demanding that all calculated triple products are of the same
sign and discarding any events where opposite sign solutions occur.

Further combinatorial issues occur with the reconstructed top in the
event. Firstly a second $b$ is always present in the opposite decay
chain and this can occasionally combine with a reconstructed W to give
a fake $t$. The opposite decay chain also can contain extra
quarks that can produce more reconstructed $t$'s. Finally, the parton
shower can sometimes radiate hard gluons that are also seen as extra
jets and further complicate the combinatorial problem. Whenever extra
$t$ quarks are found that satisfy the event kinematics, we perform the
same procedure as for combinatorial leptons. Triple products are
calculated for all reconstructed rest frames and only events, that yield the
same sign for all the reconstructed triple products, are recorded.

\subsection{Mass measurements}
\label{sec:MassMeas}

As mentioned above, we assume that the masses of all the SUSY
particles in the decay chain will be known. However, for the majority
of our equations in Eq.~(\ref{eq:MomMatEq}), we actually require the
difference between various $m^2$'s in the decay chains and not the
absolute mass. At the LHC, the established way of measuring the
SUSY spectrum is via mass end-points (see~Ref.\cite{Gjelsten:2004ki} and references
therein) and this method will measure these mass differences with high
accuracy $\mathcal{O}(1\%)$.

The on-shell mass condition for the $\neu{1}$ requires the absolute
mass scale and this should be measured at the LHC to a precision of
better than 10\%\cite{Gjelsten:2004ki}, for low mass scenarios similar
to the phenomenology presented in this paper. As an extra check on the
numerical stability of the reconstruction procedure, up to 20 GeV
absolute mass errors were tested on the absolute mass scale of the
decay chain as a conservative estimate.  This had a negligible effect
on the reconstruction efficiency and the CP-asymmetry and is therefore not
considered to be a problem. In addition new methods have been proposed
for measuring the sparticle masses from the kinematic invariants
directly
\cite{Nojiri:2003tu,Kawagoe:2004rz,Nojiri:2007pq,Cheng:2007xv,Cheng:2009fw,Casadei:2010nf}. These
methods also use the mass invariants on an event-by-event basis but
use this information to reconstruct the masses of the particles in the
decay chain. Therefore, these methods are directly measuring the
inputs we require for Eq.~(\ref{eq:chi1mass}) and
Eq.~(\ref{eq:MomMatEq}). We then use the output from these methods to
reconstruct the momentum of the $\neu{1}$ on an event-by-event basis. Reviews of all
the major mass reconstruction methods proposed for the LHC are given 
in~Refs.\cite{Barr:2010zj,Brooijmans:2010tn}.

\section{Parton level results}
\label{sec:PartResults}

In this section we analyse numerically the CP-asymmetry at the parton
level, with the inclusion of parton distribution functions, while in
Sec.~\ref{sec:HadResults} we complete a hadronic level study to estimate the effect in a realistic 
environment and
the discovery potential at the LHC. In particular, we focus on a
specific mSUGRA parameter point, Tab.~\ref{tab:Scenario}, at the
parton level before discussing more general low mass mSUGRA scenarios
for our hadronic study.

\subsection{Chosen scenario: spectrum and decay modes}
\label{sec:scenario}

We choose for this study the mSUGRA scenario shown in
Tab.~\ref{tab:Scenario} with an added CP-phase to the trilinear
coupling $\phi_{A_t}$. Although the value of the trilinear coupling is zero at the unification scale in this scenario, the renormalisation group equations generate a value of, $A_t=-391$~GeV, at the weak scale. The spectrum at the electroweak scale has been
derived using the RGE code \texttt{SPheno 2.2.3}~\cite{Porod:2003um}
and the masses of the gauginos and scalars are shown in
Tab.~\ref{tab:GaugeMasses}, Tab.~\ref{tab:SquarkMasses} and Tab.~\ref{tab:SelMasses}
respectively. Using the low energy soft SUSY breaking parameters and
the phase of the trilinear coupling $\phi_{A_t}$, we calculate the
masses and mixing of the $\st_i$'s, see Appendix~\ref{sec:StopSector} for details.

\begin{table} \renewcommand{\arraystretch}{1.3}
\begin{center}
\begin{tabular}{|c||c|c|c|c|c|}\hline
Parameter & $m_0$ & $m_{1/2}$ &  $\tan\beta$ & sign($\mu$) &  $A_0$ \\ \hline\hline
Value & 65        & 210       &   5          & +           & 0     \\ \hline
\end{tabular}
\caption{mSUGRA benchmark scenario (masses in GeV).\label{tab:Scenario}}
\end{center}
\end{table}

\begin{table}[t!] \renewcommand{\arraystretch}{1.3} 
\begin{center}
\begin{tabular}{|c||c|c|c|c||c|c||c|} \hline
Particle  & $\mneu{1}$ & $\mneu{2}$ & $\mneu{3}$ & $\mneu{4}$ & $\mcha{1}$ & $\mcha{2}$ & $m_{\tilde{g}}$ \\[0.16em]\hline\hline
Mass(GeV)  & 	77.7	&    142.4   &  305.1     &   330.3	& 140.7	    &     329.9   &  514.1     \\\hline
\end{tabular}
\caption{Masses (in GeV) of the gauginos calculated by \texttt{SPheno 2.2.3}~\cite{Porod:2003um}. \label{tab:GaugeMasses}}
\end{center}
\end{table}

\begin{table}[t!] \renewcommand{\arraystretch}{1.3} 
\begin{center}
\begin{tabular}{|c||c|c||c|c||c|c|c|c|} \hline
Particle  & $m_{\st_1}$ & $m_{\st_2}$ & $m_{\sbot_1}$ & $m_{\sbot_2}$ & $m_{\tilde{q}_{dL}}$ & $m_{\tilde{q}_{dR}}$ & $m_{\tilde{q}_{uL}}$ & $m_{\tilde{q}_{uR}}$  \\[0.16em]\hline\hline
Mass(GeV) & 	345.7   &   497.8     &    443.4      &    466.0      &    484.7	     &  465.2      	    & 
  478.7     	     &    464.9  	        \\\hline
\end{tabular}
\caption{Masses (in GeV) of the squarks calculated by
\texttt{SPheno 2.2.3}~\cite{Porod:2003um} except for the $\st_i$ which were
calculated at tree level for the phase
$\phi_{A_t}=|\frac{4}{5}\pi|$. \label{tab:SquarkMasses}}
\end{center}
\end{table}

\begin{table}[t!] \renewcommand{\arraystretch}{1.3} 
\begin{center}
\begin{tabular}{|c||c|c||c|c||} \hline
Particle  & $m_{\sl_L}$ & $m_{\sl_R}$ & $m_{\tilde{\tau}_2}$ & $m_{\tilde{\tau}_1}$ \\[0.16em]\hline\hline
Mass(GeV) & 	 163.4	  &   110.8  	    &   164.9	  &   108.0      \\\hline
\end{tabular}
\caption{Masses (in GeV) of the SUSY sleptons calculated by
\texttt{SPheno 2.2.3}~\cite{Porod:2003um}. \label{tab:SelMasses}}
\end{center}
\end{table}

For the presented analysis to work, we require the SUSY spectrum to
have the following mass hierarchy,
\begin{equation}
	 m_{\st_1} - m_t > m_{\tilde{\chi}^0_2} > m_{\tilde{\ell}^{\pm}_R} > m_{\tilde{\chi}^0_1} \label{eq:masshier},\\
\end{equation}
to allow for full momentum reconstruction. This hierarchy is often a
feature in the mSUGRA parameter space. In addition we concentrate on scenarios with a light stop as the study is statistically limited and consequently we examine cases with a large production cross section.\footnote{Since this paper was submitted, the particular parameter point, Tab.~\ref{tab:Scenario}, has been excluded \cite{Khachatryan:2011tk,daCosta:2011qk}. However, the exclusion is derived from the gluino, first and second generation squark masses but the stop masses have far weaker bounds \cite{Nakamura:2010zzi}. Therefore, if we do not restrict ourselves to a mSUGRA parameter space, the study is still valid.}

The feasibility of the method at the LHC depends heavily on the
integrated luminosity. For this reason we look closely at the predicted cross
section of the asymmetry decay chain,
\begin{equation}
\sigma=\sigma(pp \to \tilde{t}_1\tilde{t}^*_1) \times BR(\tilde{t}_1\to t \tilde{\chi}^0_2)\times BR(\tilde{\chi}^0_2 \to     \tilde{\ell}^{\pm} \ell^{\mp}) \times BR(\tilde{\ell}^{\pm} \to   \tilde{\chi}^0_1 \ell^{\pm}) \times BR(t \to q_u \bar{q}_d b),
\end{equation}
and the relevant values for our scenario are shown in
Tab.~\ref{tab:brs}. In our study we also need to identify the charge
of the $\st_1$ in the opposite decay chain and this is possible when
the decay products contain a single lepton (any number of jets are
allowed). We see that the dominant production of single leptons from
$\st_1$ decays is via the channel $\st_1 \to \cha_1^+ b$. However, as
only the right sleptons and the bino-like $\neu{1}$ are lighter than the
wino-like $\cha_1^+$, the decay of the $\cha_1^+$ is via mixing terms or Yukawa couplings and
hence the decay $BR(\cha_1^+ \to \tilde{\tau}_1^+ \nu_{\tau})$
dominates, Tab.~\ref{tab:brs}. For this reason we find that our study
is far more promising if $\tau$ identification is possible. We
compare results where $\tau$ identification (with a 40\% efficiency in the hadronic channels) has and has not been used in Sec.~\ref{sec:HadResults}.

\begin{table}[t!] \renewcommand{\arraystretch}{1.3}
\begin{center}
\begin{tabular}{|c||c|} \hline
 Parameter                     & \,\,\,\,\,\,\,\,\, Value \,\,\,\,\,\,\,\,\,  \\ \hline \hline
 $BR(\st_1 \to \neu{1} t)$    & 34.6  \\ \hline
 $BR(\st_1 \to \neu{2} t)$    & 7.5          \\ \hline 
 $BR(\st_1 \to \cha_1^+ b)$     &  50.1    \\ \hline
 $BR(\st_1 \to \cha_2^+ b)$     &  7.8   \\ \hline
 $BR(\neu{2} \to \tilde{\mu}_R^+ \mu^- / \tilde{e}_R^+ e^-)$ & 11.6   \\ \hline
 $BR(\cha_1^+ \to \tilde{\tau}_1^+  \nu_{\tau})$ & 95.1   \\ \hline \hline
 $\sigma(pp\to \st_1 \st_1^*)$ [pb] &  3.44       \\ \hline
\end{tabular}
\caption{Nominal values of the branching ratios (in $\%$) 
for various decays calculated in \texttt{Herwig++}~\cite{Bahr:2008pv,Bahr:2008tf} with phase $\phi_{A_t}=|\frac{4}{5}\pi|$. In the last row, the calculated cross section for stop pair production at the LHC with $\sqrt{s} = 14$~TeV at leading order (LO) from \texttt{Herwig++}. \label{tab:brs}}
\end{center}
\end{table}

\subsection{CP asymmetry at the parton level}\label{sec:asy-parton}

\begin{figure}[t!]
\vspace{1cm}
\begin{picture}(16,8)
 \put(-2.5,-13.5){\epsfig{file=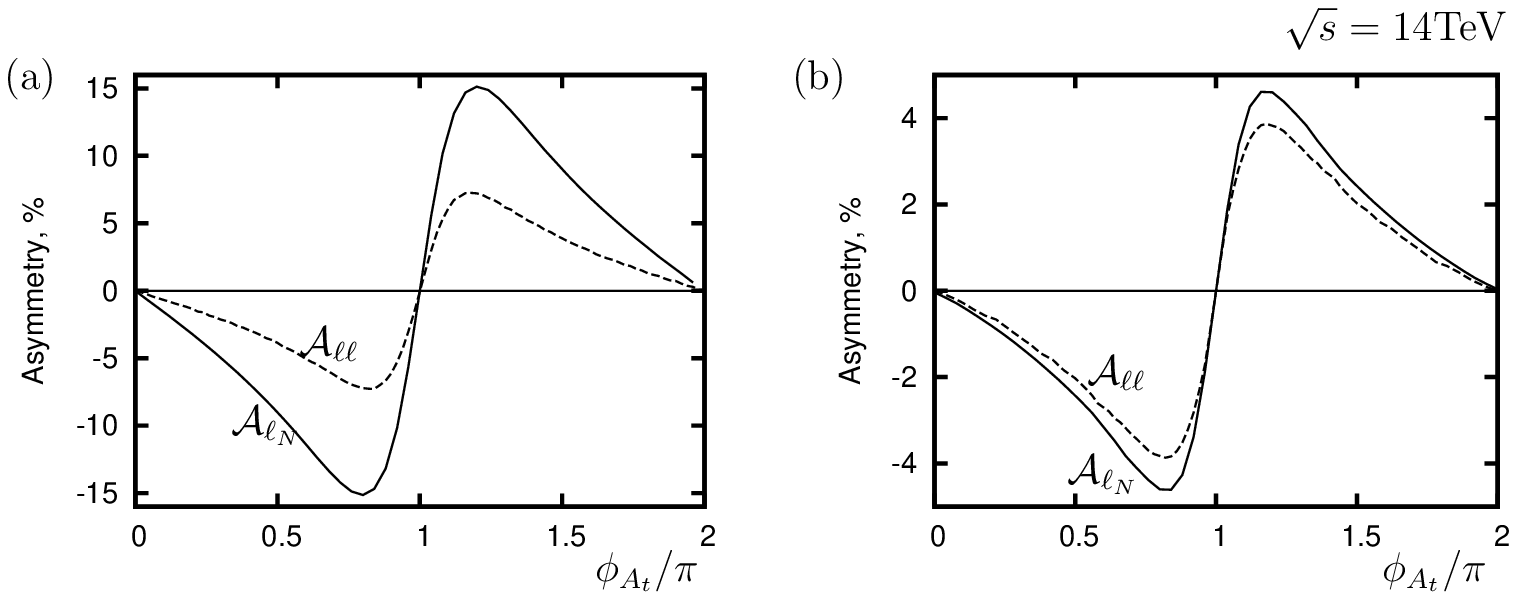,scale=1}}
\end{picture}
\vspace{-3cm}
\caption{\label{fig:PartonAsy} (a) The asymmetry
$\mathcal{A}_{\mathrm{T}}$, Eq.~(\ref{eq:Asy}), in the rest frame of
$\st_1$ as a function of $\phi_{A_t}$. (b) The asymmetry
$\mathcal{A}_{\mathrm{T}}$, Eq.~(\ref{eq:Asy}), in the laboratory
frame as a function of $\phi_{A_t}$ at the LHC at 14TeV. The solid
line is the asymmetry for the triple product $\mathcal{T}_{\ell_N}$,
Eq.~(\ref{eq:tpt}) and the dotted line is for the triple product
$\mathcal{T}_{\ell\ell}$, Eq.~(\ref{eq:tpb}).}
\end{figure}

We start by discussing the dependence of the parton
level asymmetry on $\phi_{A_t}$, Eq.~(\ref{eq:Asy}), for both the triple products
$\mathcal{T}_{\ell_N}$ and $\mathcal{T}_{\ell\ell}$,
Eqs.~(\ref{eq:tpt}), (\ref{eq:tpb}). In order to see the maximum
dependence upon $\phi_{A_t}$, we reconstruct the $\st_1$ at rest and
calculate the triple product in this frame. It should be noted that
the asymmetry is obviously a CP-odd quantity,
Fig.~\ref{fig:PartonAsy}.

We see from Fig.~\ref{fig:PartonAsy}(a) that the largest asymmetry
occurs for the triple product $\mathcal{T}_{\ell_N}$, which attains
$|\mathcal{A}_{\ell_N}|_{\mathrm{max}} \approx 15\%$ when $\phi_{A_t}
\approx 0.8\pi$. For the triple product $\mathcal{T}_{\ell\ell}$, the asymmetry
is smaller, $|\mathcal{A}_{\ell\ell}|_{\mathrm{max}} \approx 6.5\%$,
 because the \textquoteleft true' CP triple product correlation is
only partially measured, see Sec.~\ref{sec:Todd_Struc}.

If we now include the dominant production process at the LHC ($gg \to
\st_1 \overline{\st}_1$) and relevant parton distribution functions
(\texttt{MRST 2004LO} \cite{Martin:2007bv}), we see that the
asymmetries are significantly diluted,
Fig.~\ref{fig:PartonAsy}(b). The asymmetry for the triple product
$\mathcal{T}_{\ell_N}$, drops from $|\mathcal{A}_{\ell_N}|_{\mathrm{max}} \approx
15\%$ to $|\mathcal{A}_{\ell_N}|_{\mathrm{max}} \approx 4.5\%$ and the
reduction is due to the boosted frame of the produced $\st_1$ as
discussed in Sec.~\ref{sec:Todd_Struc}. For the triple product
$\mathcal{T}_{\ell\ell}$, the reduction in the asymmetry is far less, from
$|\mathcal{A}_{\ell\ell}|_{\mathrm{max}} \approx 6.5\%$ to
$|\mathcal{A}_{\ell\ell}|_{\mathrm{max}} \approx 3.8\%$. This is because the
triple product, relies on the $\ell_F$ being correlated with the
$\tilde{\ell}$ by the intrinsic boost of the $\neu{2}$, $\tilde{\ell}$
system which already has a boost, even when the $\st_1$ is at rest. As
the $\st_1$ becomes boosted, the boost of the $\neu{2}$,
$\tilde{\ell}$ system becomes proportionally less so, as the momentum
of the $\st_1$ is distributed throughout the decay chain. The
difference in the dilution of the two asymmetries with $\st$ momentum
can be seen in Fig.~\ref{fig:MomPlot}.

\section{Hadron level results}
\label{sec:HadResults}

In order to estimate
the potential for observing
CP-violating effects in $\st_1$ decays at the LHC more realistically, we perform the
analysis at the hadronic level. We use the
\texttt{Herwig++}~\cite{Bahr:2008pv,Bahr:2008tf} event generator to
calculate all the matrix elements in the process, the initial hard
interaction, the subsequent SUSY particle decays, the parton shower
and the hadronisation. An important feature of \texttt{Herwig++} is
that it calculates the spin correlations \cite{Richardson:2001df} in the SUSY cascade decay and
allows the input of complex mixing matrices. Consequently, the triple
product CP-asymmetry can be automatically calculated within
\texttt{Herwig++}.

We also include both Standard Model and SUSY backgrounds in the analysis to understand how well $\st_1\st^*_1$ production can be isolated at the LHC. We find that after applying basic signal identification cuts and the more complicated momentum reconstruction, virtually no Standard Model background contributes, Sec.~\ref{sec:sm-at-lhc}. However, the SUSY background presents more of a challenge and new cuts have to be introduced to improve the signal to background ratio, Sec.~\ref{sec:susy-process-at-lhc}. Even after cuts, the SUSY background can remain problematic but if the dominant contributions are known, the backgrounds can be partially subtracted.

\subsection{Cuts used and signal identification}


The hadronic analysis of the
produced events has been performed 
within the program
\texttt{Rivet}~\cite{Buckley:2010ar,Waugh:2006ip}. We used the
anti-$\mathrm{k_t}$ \cite{Cacciari:2005hq,Cacciari:2008gp} jet
algorithm with R=0.5 and applied the following acceptance cuts,

\begin{itemize}
\item 
	$p_{T\ell_i} > 10 \mathrm{GeV}$,
\item 
	$p_{Tj_i} > 20 \mathrm{GeV}$,
\item
	invariant mass of opposite sign same flavour (OSSF) leptons: $M_{\ell^+\ell^-} > 10 \mathrm{GeV}$,
\item 
	$|\eta_{\ell_i}| < 2.5$,
\item 
	$|\eta_{j_i}| < 3.5$,
\item
	lepton jet isolation, $\Delta$R = 0.5,
\item
	$b$-tag efficiency = 60\% \cite{Aad:2009wy},
\item
	hadronic $\tau$-tag efficiency (whenever used) = 40\% \cite{Aad:2009wy}.
\end{itemize}

To identify the events we demand three charged leptons in
the final state, so that we can correctly identify the charge of each
$\st_1$ produced in the event, c.f.~Sec.~\ref{sec:Todd_Struc}. In addition,
we demand that a pair of these leptons are OSSF as is the case for light leptons
from $\neu{2}$ decay. Whenever a $\st_1$ decays in our scenarios a
$b$ is produced and therefore we require at least one $b$-tag in the
final state (in principle we could require 2 $b$-tags including the
opposite decay chain but we lose~40\% of events due to b-tagging
efficiency). On top of the $b$ we require at least 2 more jets to be
found in the final state so the full reconstruction of the $t$ is
possible. As all of our triple products and reconstructed momenta
need a $t$, we require at least one hadronic $t$ to be
reconstructed. For this procedure, we first demand that 2 jets (not
$b$'s) reconstruct a $W^{\pm}$ ($70\mathrm{GeV} < M_{jj} <
90\mathrm{GeV}$). We then impose that a reconstructed $W^{\pm}$ and
one $b$ jet reconstruct a $t$ ($150\mathrm{GeV} < M_{W^{\pm}b} <
190\mathrm{GeV}$).


Once these cuts have been passed we then perform the kinematical
reconstruction shown in Sec.~\ref{sec:MomRec} with any $t$'s and OSSF
leptons found in the final state. If the particles satisfy the
kinematic constraints Eqs.~(\ref{eq:chi1mass})-(\ref{eq:stopmass}), we
will have at least two different solutions on an event-by-event basis
for the momentum of the
$\neu{1}$. For each solution, the relevant rest frame triple product
is calculated and only if all the signs of the triple products agree 
the event is accepted.

\subsection{Standard model background}
\label{sec:sm-at-lhc}

The following standard model backgrounds were produced with
\texttt{Herwig++}: $t \overline{t}$, Drell-Yan (via $\gamma$ and $Z$), $W$+jet, $Z$+jet, $WW$, $WZ$, $ZZ$ and $W\gamma$. In addition, we generate $t \overline{t} \ell^+ \ell^-$ events
with \texttt{MadGraph} \cite{Alwall:2007st} and then use
\texttt{Herwig++} to perform the parton shower and hadronisation. We
find that after we produce an equivalent luminosity of 500~fb$^{-1}$, the only background to pass the event selection is $t
\overline{t} \ell^+ \ell^-$ with the very low rate of 0.03
events/fb$^{-1}$ after kinematical reconstruction. This corresponds to only
$\approx 1\%$ of the signal process for our particular scenario.

Although the above result is encouraging, it must be stated that our
analysis contains no jets mis-identified as leptons. As the dominant standard
model contributions produced by \texttt{Herwig++} only contain two hard leptons in the initial process, the lack of a trilepton
signal is not  surprising. However, we do not expect major problems
from standard model backgrounds if we limit the study to leptons from
the first and second generation. $t \overline{t}$ can be expected to
provide the largest background when both $W^{\pm}$ decay leptonically
and an extra lepton is produced from a $b$ or a mis-identified
jet. Even when this occurs though, we still require an additional
two hard jets in the event that have to combine with a $b$ to form a
$t$. Moreover, the final state then has to fulfil the reconstructed particular
kinematics of our signal and finally all the calculated triple
products have to agree.

To improve the statistical significance of our analysis, we also
investigated the possibility of using $\tau$-tagging in the opposite
decay chain to that of our signal. In this analysis, we now change the
original trilepton signal to a first or second lepton OSSF and
additional hadronic $\tau$. The mis-identification of a jet for a
$\tau$ is much higher than for the other leptons and the standard
model backgrounds may now become an issue \cite{Aad:2009wy}. However, this analysis is postponed to future studies.


\begin{figure}[t!]
\vspace{1cm}
\begin{picture}(16,8)
 \put(-2.5,-13.5){\epsfig{file=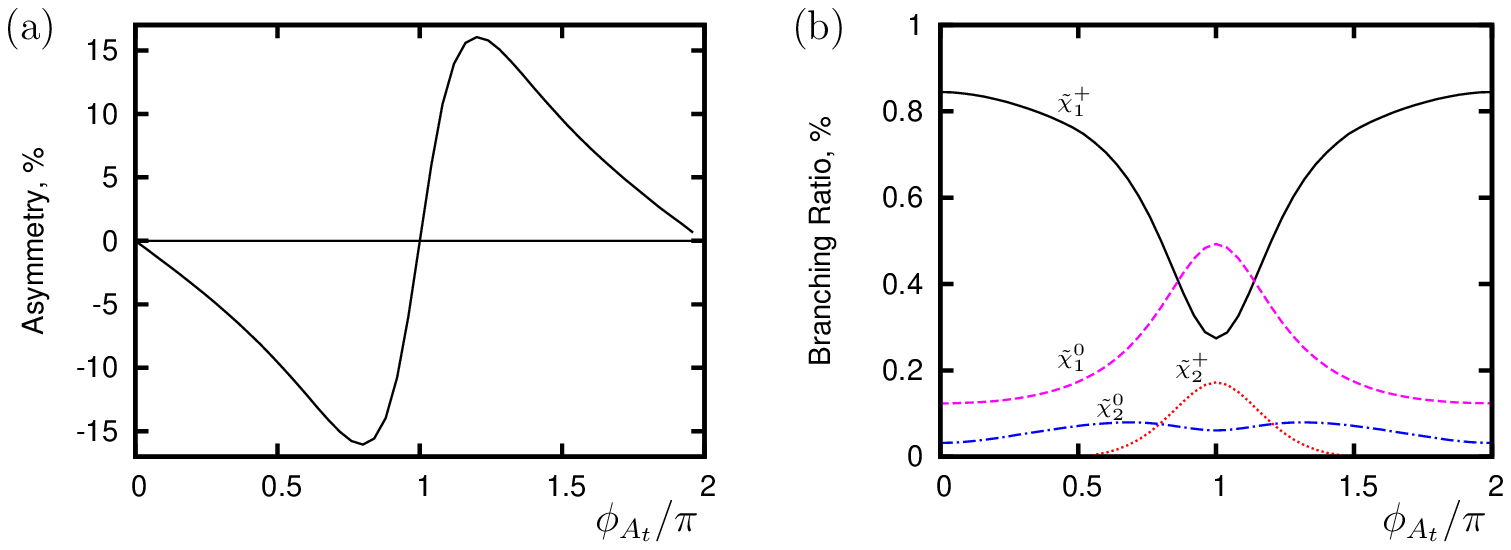,scale=1}}
\end{picture}
\vspace{-3cm}
\caption{\label{fig:NeutChaAsy} (a) The asymmetry
$\mathcal{A}_{\ell_N}$, Eq.~(\ref{eq:Asy}), for the decay chain
shown in Eq.~(\ref{eq:decaysig})-(\ref{eq:oppsig}) as a function of
$\phi_{A_t}$ at the hadronic level after momentum reconstruction has
been performed. (b) The branching ratios: $\tilde{t}_1
\to \tilde{\chi}^+_1 b$  (black solid), $\tilde{t}_1 \to
\tilde{\chi}^+_2 b$ (red dotted),  $\tilde{t}_1 \to
\tilde{\chi}^0_1 t$ (purple dashed), $\tilde{t}_1 \to
\tilde{\chi}^0_2 t$  (blue dash-dot).}
\end{figure}

\subsection{Stop Production}
\label{sec:stop-prod-at-lhc}

We begin by studying $\st_1\st_1^*$ production along with the
following decay chain,
\begin{eqnarray}
	\tilde{t}_1 & \to & \tilde{\chi}^0_2 t \to \neu{1} e^+ e^- j_{u} j_{\bar{d}}
	b, \label{eq:decaysig}\\ \tilde{t}_1^* & \to & \neu{1}
	\overline{t} \to \neu{1} \mu^- \bar{\nu}_{\mu} b.  \label{eq:oppsig}
\end{eqnarray}
to test the momentum reconstruction procedure. The above decay chain
is the cleanest signal process from a combinatorial point of view. We find a reconstruction efficiency of $\approx 5\%$ for this
particular topology after cuts and the requirement for same sign triple products. The decay chain Eq.~(\ref{eq:oppsig}) has a
single lepton in the final state allowing us to tag the charge of both
the $\st_1$ and $\st_1^*$ in the process.

For the CP-asymmetry, we now concentrate purely on the triple product
$\mathcal{T}_{\ell_N}$, Eq.~(\ref{eq:tpt}), calculated in the reconstructed
rest frame of the $\st_1$, as this is the observable with high
significance at the LHC. Fig.~\ref{fig:NeutChaAsy}(a) shows that there
is virtually no dilution when we move to the hadronic level and the maximal
asymmetry stays at $|\mathcal{A}_{\ell_N}|_{\mathrm{max}} \approx 15\%$. In
fact, the hadronic level reconstruction does induce a degree of
dilution, $\approx 1.5\%$ but this is cancelled by our procedure of
removing opposite sign triple products which enhances the asymmetry by
a similar amount, c.f.~Sec.~\ref{sec:MomRecPrac}.

In order to estimate whether it is possible to observe a CP-asymmetry in $\st_1$ decays at the LHC, we need to calculate the statistical significance of any result. We assume that $N_{\mathcal{T}_+}\,(N_{\mathcal{T}_-})$, the numbers of events where $\mathcal{T}$ is positive (negative) as in Eq.~(\ref{eq:Asy}), are binomially distributed, giving the following statistical 
error~\cite{Desch:2006xp},
\begin{equation}
  \label{eq:asymmerror}
  \Delta(\mathcal{A}_T)^{\rm stat}=2\sqrt{\epsilon(1-\epsilon)/N}\,,
\end{equation}
where $\epsilon=N_{\mathcal{T}_+}/(N_{\mathcal{T}_+}+N_{\mathcal{T}_-})=\frac 12
(1+\mathcal{A}_T)$, and $N = N_{\mathcal{T}_+}+N_{\mathcal{T}_-}$ is the total number of events. Eq.(\ref{eq:asymmerror}) can be rearranged to
give the required number of events for a desired significance.

\begin{table}[t!] \renewcommand{\arraystretch}{1.3}
\begin{center}
\begin{tabular}{|c||c|c|} \hline
        & $\st_1\st_1^*$ & $\tilde{g},\tilde{q}$ \\ \hline \hline
  Herwig++ LO (pb$^{-1}$)    & $3.44$ & $75.8$  \\ \hline 
 Prospino LO (pb$^{-1}$)    & $3.34^{+1.15}_{-0.8}$ & $76.7^{+24.8}_{-17.3}$  \\ \hline 
 Prospino NLO (pb$^{-1}$)   & $5.04^{+1.19}_{-0.92}$ & $99.5^{+7.7}_{-9.6}$    \\ \hline  
\end{tabular}
\caption{Cross section at the LHC with $\sqrt{s} = 14$~TeV production channel $\st_1\st_1^*$ and coloured SUSY production for both leading order (LO) and next-to-leading order (NLO). All cross sections were calculated using \texttt{Herwig++}~\cite{Bahr:2008pv,Bahr:2008tf} or \texttt{Prospino}~\cite{Beenakker:1996ed,Beenakker:1996ch,Beenakker:1997ut}. The errors indicated next to the \texttt{Prospino} cross sections relate to varying the factorisation and renormalisation scales from $0.5m_{\st_1} \to 2m_{\st_1}$. \label{tab:Sig_Prosp}}
\end{center}
\end{table}

The total cross section used to calculate the statistical significance
of any result in this paper has been calculated using
\texttt{Herwig++} at the leading order (LO) for consistency. However,
next-to-leading order production cross sections are available and have
been calculated using
\texttt{Prospino}~\cite{Beenakker:1996ed,Beenakker:1996ch,Beenakker:1997ut},
cf.\ Tab.~\ref{tab:Sig_Prosp}. We see that in general the cross
sections at NLO are higher than those at LO suggesting that the
effective luminosity at the LHC will be more optimistic than those
shown in the following results. In addition, the factorisation and
renormalisation scale uncertainties are shown that indicate an estimate of
the underlying theoretical uncertainty.

Due to the phase dependence of both the $\st_1$ branching ratios, see
Fig.~\ref{fig:NeutChaAsy}(b), and production cross section, the
statistical significance for different values of $\phi_{A_t}$ cannot
be trivially extrapolated. The total number of events observed will be
an interplay between the branching ratios and the production cross
section. However, in the case of branching ratios, each of the decays,
$\tilde{t}_1 \to \tilde{\chi}^+_1 b$, $\tilde{t}_1 \to
\tilde{\chi}^+_2 b$ and $\tilde{t}_1 \to \tilde{\chi}^0_1 t$ has a
different reconstruction efficiency and asymmetry dilution that needs
to be calculated. For example, we see from
Fig.~\ref{fig:NeutChaAsy}(b) that the branching ratio for the decay
$\tilde{t}_1 \to \tilde{\chi}^+_2 b$ increases noticeably as we vary $\phi_{A_t}$
from $\phi_{A_t}=0$ to $\phi_{A_t}=|\pi|$ due to this decay becoming
kinematically more favourable. The $\tilde{\chi}^+_2$ has a large
number of final states with no lepton however, so consequently the
number of signal events decreases. Also, the $\tilde{\chi}^+_2$ decays
generally contain extra jets that make the reconstruction of the event
more difficult and thus reduce the efficiency of this channel.

\begin{figure}[t!]
\vspace{1cm}
\begin{picture}(16,7)
 \put(-2.5,-13.5){\epsfig{file=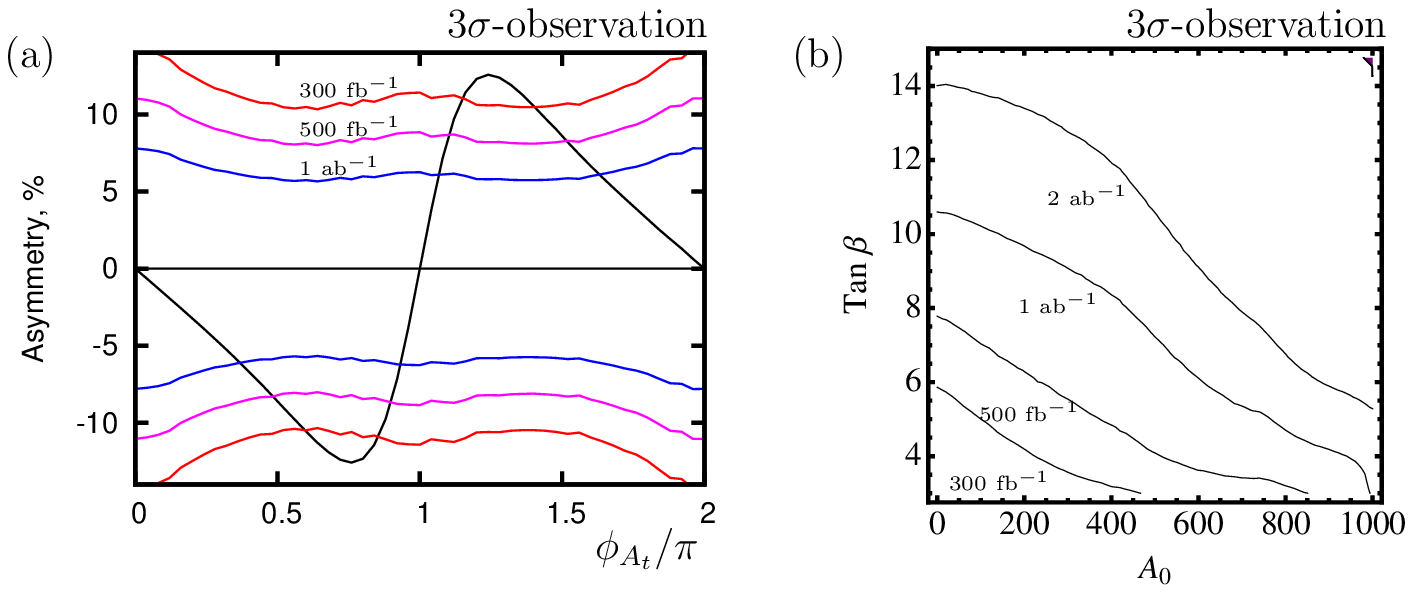,scale=1}}
\end{picture}
\vspace{-3cm}
\caption{\label{fig:NoCutsA0} Pure $\st_1\st_1^*$ production, all decay channels included, see Tab.~\ref{tab:brs} for branching ratios for the specific parameter point and Fig.~\ref{fig:NeutChaAsy} for how these alter with $\phi_{A_t}$. $\tau$ tagging is included in both plots. (a) Asymmetry, $\mathcal{A}_{\ell_N}$, at reference point with 3$\sigma$-luminosity lines shown. (b) Minimum luminosity required for 3$\sigma$-discovery in tan$\beta, A_0$ plane (at the unification scale) when asymmetry, $\mathcal{A}_{\ell_N}$, is maximal.}
\end{figure}

Figure~\ref{fig:NoCutsA0}(a) shows the asymmetry when all $\st_1$ decay
channels are considered and an estimate of the amount of luminosity
required for a 3$\sigma$-observation (statistical errors only) of a non-zero asymmetry for pure
$\st_1 \st^*_1$ production at the LHC. We can see that the asymmetry
is slightly diluted when all $\st_1$ decay modes are included from
$|\mathcal{A}_{\ell_N}|_{\mathrm{max}} \approx 15\%$ to
$|\mathcal{A}_{\ell_N}|_{\mathrm{max}} \approx 12.5\%$. The dilution is due
to reconstructed events that are not originating from the signal process,
Eq.~(\ref{eq:stopdecay}). These events have no
overall asymmetry and therefore simply dilute the signal. The
horizontal lines show the estimate of the required luminosity required
to see a certain asymmetry; an asymmetry can be seen at the 3$\sigma$
level where the asymmetry curve in Fig.~\ref{fig:NoCutsA0}(a) lies
outside the luminosity band. The luminosity bands are not flat because
as discussed before, both the branching ratios and production cross
section of the $\st_1$ vary with the phase $\phi_{A_t}$. We can see
that in our scenario for pure $\st_1 \st^*_1$ production, we expect
a sensitivity  for $0.5\pi < \phi_{A_t}(\mathrm{mod}\; \pi) < 0.9\pi$ with 500 fb$^{-1}$. With a combined analysis of both ATLAS and CMS data, this luminosity can be expected to be reached in the early 2020's if the LHC operates as is currently planned \cite{LHC1}.

\begin{figure}[t!]
\vspace{1cm}
\begin{picture}(16,7)
   \put(-2.5,-13.5){\epsfig{file=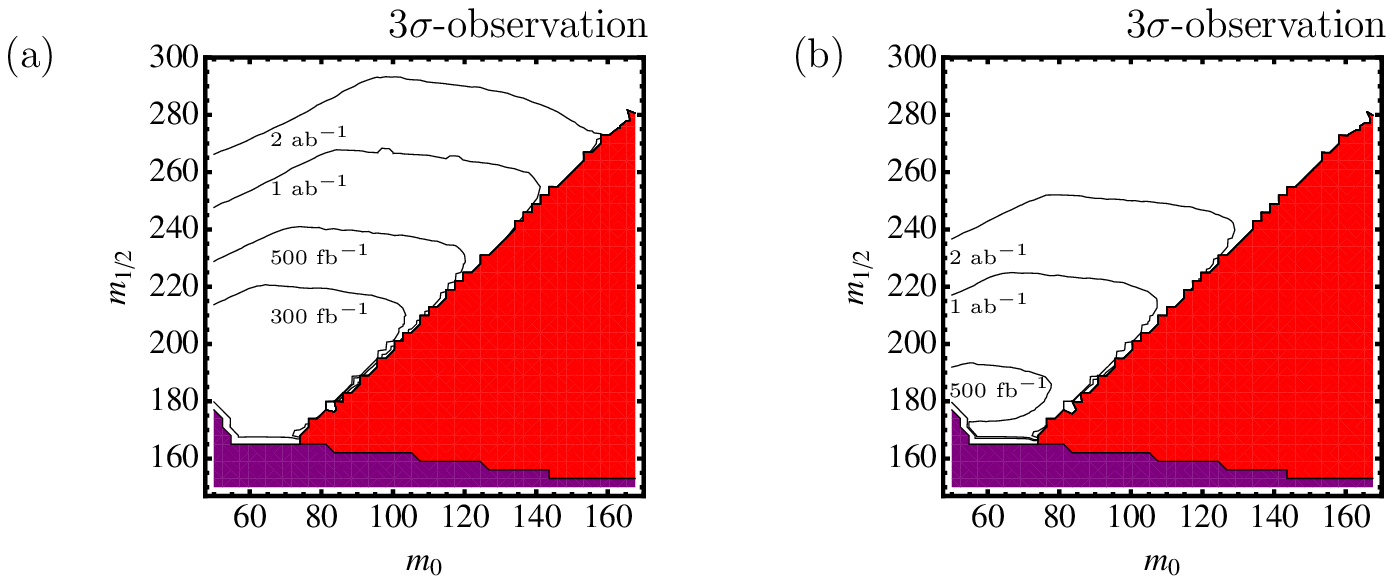,scale=1}}
\end{picture}
\vspace{-3cm}
\caption{\label{fig:NoCutsM0} Minimum luminosity required for 3$\sigma$-discovery in $m_0, m_{1/2}$ plane (at unification scale) when asymmetry,  $\mathcal{A}_{\ell_N}$, is maximal. Pure $\st_1 \st_1^*$ production, all decay channels included, see Tab.~\ref{tab:brs} for branching ratios for the specific parameter point and Fig.~\ref{fig:NeutChaAsy} for how these alter with $\phi_{A_t}$.  Purple area is ruled out by LEP direct detection \cite{Amsler:2008zzb} and red area has no two body decay $\tilde{\chi}^0_2 \to \tilde{\ell}^{\pm} \ell^{\mp}$. (a) With $\tau$ tagging. (b) Without $\tau$ tagging.}
\end{figure} 

We can see the effect of varying the mSUGRA parameters
$\mathrm{tan}\beta$ and $A_0$ in Fig.~\ref{fig:NoCutsA0}(b). It is
shown that as the value of either $\mathrm{tan}\beta$ or $A_0$ is
increased, we require more luminosity to see a statistically
significant observation even with maximum asymmetry. An increase in
$\mathrm{tan}\beta$ decreases the sensitivity because the branching
ratio $\tilde{\chi}^0_2 \to \tilde{\ell}^{\pm} \ell^{\mp}$ is
reduced. The reduction is due to $\tilde{\tau}$'s becoming more mixed
which increases the left handed component in the lighter
$\tilde{\tau}$. Therefore, the $\tilde{\tau}_1$ couples more strongly
to the predominantly wino $\neu{2}$ and begins to dominate this decay
channel at the expense of the signal process. A rise in $A_0$ decreases
sensitivity mainly because the CP-asymmetry is reduced. The
reason is that after RGE running, an increase in $A_0$ reduces the
magnitude of the trilinear coupling $A_t$ that contains the phase,
$\phi_{A_t}$ that we are interested in. Hence the CP effects are
reduced. 

Similarly, Fig.~\ref{fig:NoCutsM0}(a) shows the effect of varying the
mSUGRA parameters $m_0$ and $m_{1/2}$ on the minimum luminosity
required for an observation of CP effects. We note as a general trend 
that as $m_{1/2}$ is increasing, we need more luminosity to observe the
CP-violating triple products. This is due to the increase in $\st_1$
mass which reduces the production cross section for $\st_1
\st_1^*$. If we increase $m_0$ we see that a large area of the
parameter space has no two body decay $\tilde{\chi}^0_2 \to
\tilde{\ell}^{\pm} \ell^{\mp}$ as $m_{\tilde{\ell}^{\pm}} >
m_{\tilde{\chi}^0_2}$.

Fig.~\ref{fig:NoCutsM0}(b) indicates the effect of having no hadronic $\tau$-tagging for the decay $\tilde{\chi}^+_1 \to \tilde{\tau}^+_1 \nu_{\tau}$. The $\tau$ final state dominates the $\tilde{\chi}^+_1$ decay which in turn is the dominant product of the $\st_1$ in low mass mSUGRA scenarios, Tab.~\ref{tab:brs}. As stated in the beginning of Sec.~\ref{sec:HadResults} we assume a 40\% $\tau$-tagging efficiency and without this we lose approximately a factor of 2 in effective luminosity for our signal process.

\subsection{Impact of momentum reconstruction on SUSY background separation}
\label{sec:susy-process-at-lhc}

All of the previous section's results have assumed that the $\st_1
\st_1^*$ process can be isolated effectively. However, in the mSUGRA
scenarios investigated many other SUSY particles will be
produced. Table~\ref{tab:Sig_Back} shows that the total production
cross section for SUSY is $\approx25$ times greater than for
$\st_1\st_1^*$ production and we can therefore expect sizable
backgrounds. We can also expect that the vast majority of the SUSY
background processes will have no other spin correlated CP-sensitive triple product with
the same final state and will therefore just act as a dilution to the
CP-asymmetry by contributing to the denominator of Eq.~(\ref{eq:Asy}).

Table~\ref{tab:Sig_Back} shows that after the initial event selection
and top reconstruction, the SUSY background is still $\approx10$ times
larger than the signal process. Note that if we apply the
kinematical reconstruction to these events we see that we
substantially reduce the background to be only $\approx3$ times larger.

\begin{table}[t!] \renewcommand{\arraystretch}{1.3}
\begin{center}
\begin{tabular}{|c||c|c|c|} \hline
        & $\st_1\st_1^*$ & SUSY & $\st_1\st_1^*$ Signal / SUSY Background \\ \hline \hline
 Cross Section (pb$^{-1}$)    & 3.44 & 80.1 & \\ \hline \hline
 Events with 500 fb$^{-1}$    & 1.7$\times10^{6}$ & 4$\times10^{7}$ &      \\ \hline 
 Events with 500 fb$^{-1}$    & 32389 & 410735 &  0.079      \\  
	Initial selection & & &\\ \hline
 Events with 500 fb$^{-1}$    & 7117 & 64729 &  0.11      \\ 
	Top Reconstruction & & &\\ \hline
 Events with 500 fb$^{-1}$      &  1213  & 3759 &  0.32 \\ 
        Kinematic Reconstruction & & &\\ \hline
 Events with 500 fb$^{-1}$   &  901  &  967  & 0.93 \\ 
       Extra SUSY cuts  & & &\\ \hline  
\end{tabular}
\caption{Cross section, number of events and signal to background
ratio at the LHC with $\sqrt{s} = 14$~TeV at LO for both the
production channel $\st_1\st_1^*$ and inclusive SUSY production. All
cross sections were calculated using
\texttt{Herwig++}~\cite{Bahr:2008pv,Bahr:2008tf}. \label{tab:Sig_Back}}
\end{center}
\end{table}

In order to observe CP-violating effects in $\st_1
\st^*_1$ production at the LHC, however, the signal to background ratio
may still be too high and consequently we need further cuts to isolate the
signal process. We notice that in mSUGRA scenarios, the largest
background comes from $\tilde{g}$ production followed by the dominant
decay to either sbottom, $\tilde{g} \to \tilde{b}_i b$ with a branching ratio of
$\approx 30\%$. The $\tilde{b}_i$ decays dominantly to $\neu{2} b$ or
$\cha^+_1 t$ which leads to
a very similar final state as the signal process
when combined with the opposite decay chain. The difference between
the SUSY background and the $\st_1$'s is that the $\tilde{g}$ and
first and second generation $\tilde{q}$ have a higher mass. In
addition, a gluino has in general one more decay vertex in the
cascade decay producing another hard jet. These two factors mean that
the average $p_T$ of the particles produced in the event will be
higher and the number of jets will be greater, thus we can use these
characteristics to discriminate the signal from the background. Hence
we cut on the number of jets reconstructed in an event,
\begin{eqnarray}
  \mathrm{Number \;of \;jets} < 6.
\end{eqnarray}
For the $p_T$ cuts, we have,
\begin{eqnarray}
  p_T(\mathrm{Hardest\; Jet}) & < & 200 \;\mathrm{GeV}, \\
  p_T(\mathrm{2nd\; Jet}) & < & 130 \;\mathrm{GeV}, \\ 
  p_T(\mathrm{3rd\; Jet}) & < & 80 \;\mathrm{GeV} \mathrm{\;(if \;applicable)},\\
  p_T(\mathrm{Any\; } b\mathrm{\; Jet}) & < & 150 \;\mathrm{GeV}, \\
  p_T(\mathrm{Any\; Lepton}) & < & 100 \;\mathrm{GeV}.
\end{eqnarray}


Table~\ref{tab:Sig_Back} shows that after all these cuts are performed
the signal to background ratio improves significantly and we now have
roughly the same number of signal and background events in the sample.

\begin{figure}[t!]
\vspace{1cm}
\begin{picture}(16,7)
 \put(-2.5,-13.5){\epsfig{file=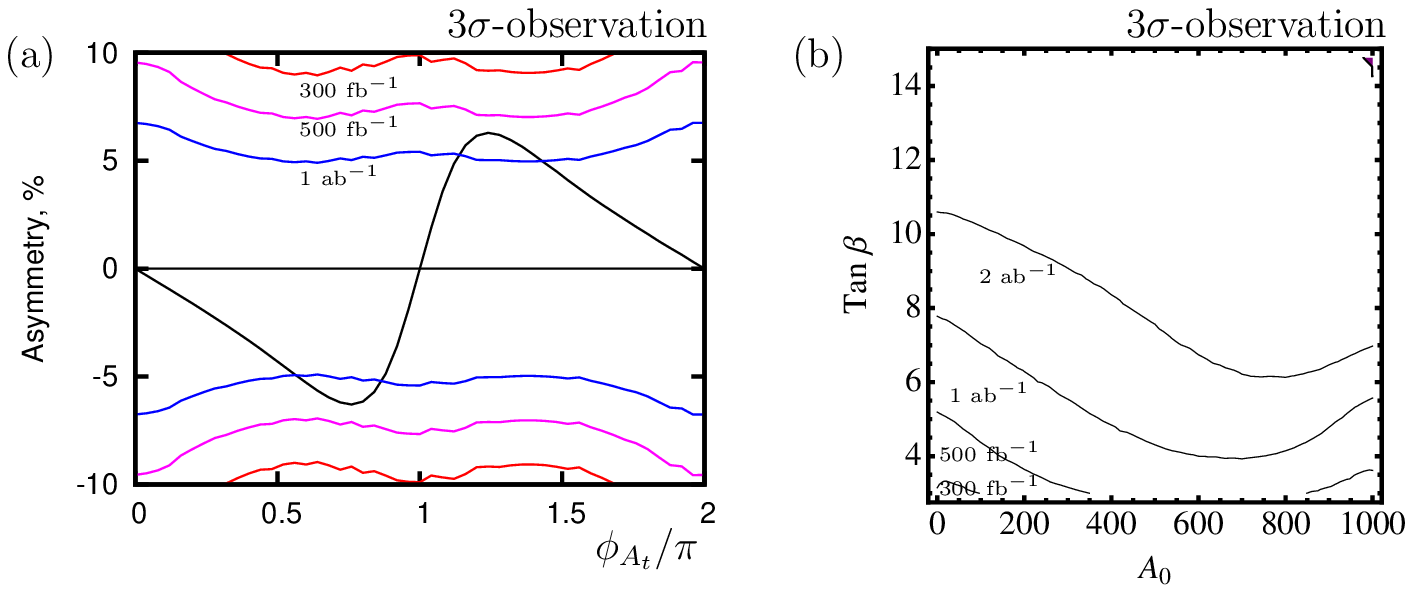,scale=1}}
\end{picture}
\vspace{-3cm}
\caption{\label{fig:SUSYA0} General SUSY production for the asymmetry $\mathcal{A}_{\ell_N}$. $\tau$ tagging is included in both plots. (a) Asymmetry, $\mathcal{A}_{\ell_N}$, at reference point with 3$\sigma$-luminosity lines shown. (b) Minimum luminosity required for 3$\sigma$-discovery in tan$\beta, A_0$ plane (at unification scale) when asymmetry, $\mathcal{A}_{\ell_N}$, is maximal.}
\end{figure}

\begin{figure}[t!]
\vspace{1cm}
\begin{picture}(16,7)
  \put(-2.5,-13.5){\epsfig{file=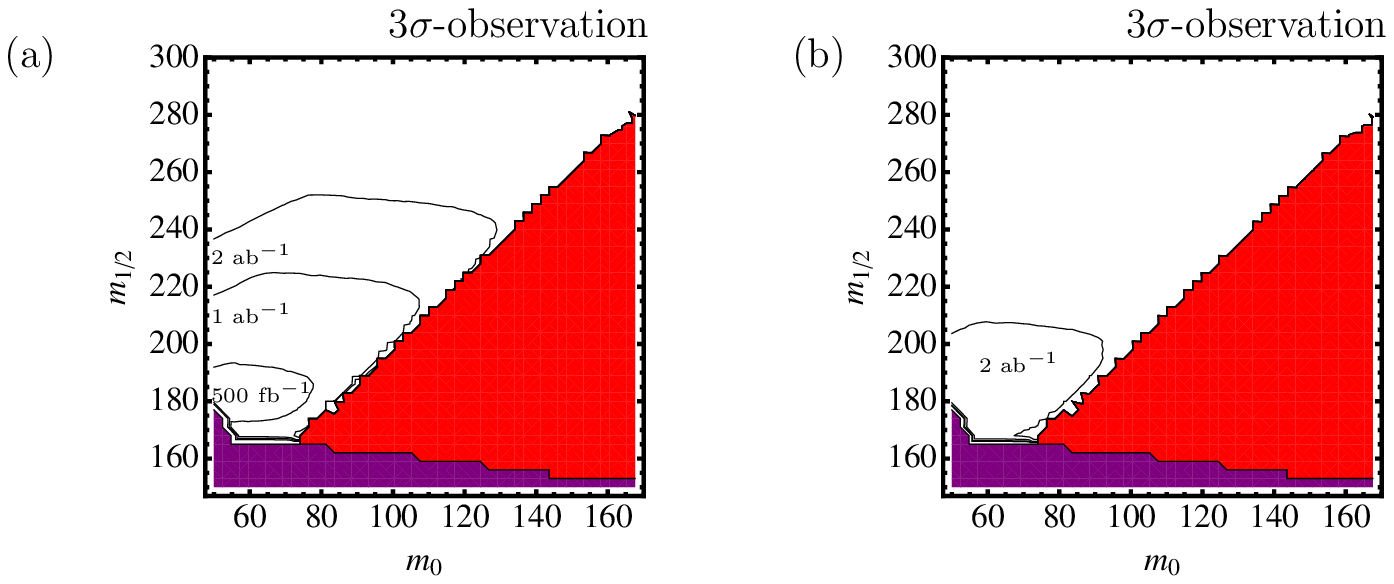,scale=1}}
\end{picture}
\vspace{-3cm}
\caption{\label{fig:SUSYM0} General SUSY production for the asymmetry $\mathcal{A}_{\ell_N}$. Minimum luminosity required for 3$\sigma$-discovery in $m_0, m_{1/2}$ plane (at unification scale) when asymmetry, $\mathcal{A}_{\ell_N}$, is maximal. Purple area is ruled out by LEP direct detection \cite{Amsler:2008zzb} and red area has no two body decay $\tilde{\chi}^0_2 \to \tilde{\ell}^{\pm} \ell^{\mp}$. (a) With $\tau$ tagging. (b) Without $\tau$ tagging. }
\end{figure}

If we now re-evaluate the luminosity plots with the SUSY background
included, Fig.~\ref{fig:SUSYA0},\ref{fig:SUSYM0}, we see that more
luminosity is now required to observe a statistically significant
effect. Due to the background dilution of the asymmetry, we now have
$|\mathcal{A}_{\ell_N}|_{\mathrm{max}} \approx 6.5\%$ for our scenario
Fig.~\ref{fig:SUSYA0}. Consequently we are now only sensitive to
phases between $0.6\pi < \phi_{A_t}(\mathrm{mod}\; \pi) < 0.85\pi$ with 1 ab$^{-1}$ of
data. If we look at the tan$\beta$, $A_0$ contour plot we see that
sensitivity at the LHC for 1 ab$^{-1}$ is only possible for small
values of tan$\beta$. A luminosity of 1 ab$^{-1}$ would probably require an upgrade to the High-Luminosity LHC (HL-LHC) \cite{LHC2}.

However, we would like to emphasise that it may be possible to substantially improve the statistical
significance of an asymmetry measurement and return to close to the
significance achieved when looking at a purely $\st_1 \st_1^*$ process,
even with the same SUSY background. Namely, via measuring
the SUSY spectra (in particular the $\tilde{g}$ and $\tilde{b}$) a good
estimate of the background should be possible. The background events
can then be subtracted from the denominator of the asymmetry,
Eq.~(\ref{eq:Asy}), to give the true value of the asymmetry. Thus, the
statistical significance should be much improved.

We would also like to remind the reader that this subtraction only becomes reliable if the signal to background ratio is good enough, otherwise the signal is swamped by statistical fluctuations. Thus the momentum reconstruction procedure is vital since it significantly reduces the backgrounds that are present.

Similarly, a more constrained area of observability is seen in the
$m_0$, $m_{1/2}$ plane, Fig.~\ref{fig:SUSYM0}(a). With 1 ab$^{-1}$ of
data, our study suggests that only if $m_{1/2}<220$ GeV will it be
possible to observe a CP-phase in the stop sector. Again, we see the
importance of $\tau$-tagging to our study from the difference between
Fig.~\ref{fig:SUSYM0}(a) and Fig.~\ref{fig:SUSYM0}(b). If
$\tau$-tagging is not used in the study, no CP-violation in the
$\st_1$ sector can be observed with 1 ab$^{-1}$ of data.

\subsection{Open experimental issues}

Although the presented study was completed at the hadronic level, a full detector simulation should be completed to confirm the conclusions of this paper. The most obvious experimental issue that could affect our results is the finite momentum resolution of the detector for both jets and leptons when performing momentum reconstruction. However, the resolution was tested with regards to momentum reconstruction in~Ref.\cite{MoortgatPick:2009jy} with a significantly more complicated final state and it was found to have only a small effect.

In terms of background suppression the mis-tagging of various objects could increase both the standard model and SUSY background. For the standard model background, the most obvious example is the $t \overline{t}$ process generating a trilepton signal \cite{Aad:2009wy}. The process requires a jet to be mistagged as a lepton, which is not investigated in this study. The suitability of hadronic $\tau$-tagging in the study also needs to be investigated thoroughly as these are expected to have significant mis-identification rates \cite{Aad:2009wy}. However, this is beyond the scope of this theoretical study.


\section{Conclusions}

In this paper we have investigated the problem of discovering
CP-violating effects at the Large Hadron Collider. We studied
$\st_1\st_1^*$ production and subsequent two body decays. Triple
product correlations can be formed from the final state particles that
are sensitive to the presence of complex phases in the model. Since
triple products depend crucially on spin correlations and are
therefore sensitive to CP-odd observables, they have been included
both in the analytical calculation and the event generation, that has
been performed using \texttt{Herwig++ 2.3.2}. The process of special
interest in our case was the $\st_1$ decay into $t$ and
$\tilde{\chi}_2^0$ followed by two, 2-body leptonic decays. For this
decay in our mSUGRA scenario one can expect an asymmetry in the triple
product distribution of up to 15\% when calculated in the rest frame
of the decaying neutralino. The source of the CP violation in our case
was the phase of the trilinear coupling $A_t$ that attains a value of $\phi_{A_t}\sim0.8$ when the asymmetry is maximum in our scenario.

Due to the hadronic experimental environment of the LHC, precise
measurements will be a challenge both from experimental and
theoretical point of view. The rest frame CP-odd asymmetry is diluted
by the high boosts of the produced particles and this makes an
observation difficult. We studied the impact of momentum reconstruction
of invisible LSPs to get access to the rest frame of the
$\st_1$. Using a set of invariant kinematic conditions we showed that
it is possible to fully reconstruct the production and decay process
on an event-by-event basis. The reconstruction was performed on events
including the parton shower and hadronisation. Having fully
reconstructed events we are able to boost particle momenta back to the
rest frame of the $\st_1$ and the maximum asymmetry is recovered to
15\%. In addition, momentum reconstruction leads to a significant increase in the signal background ratio and thus is very important in attempting to isolate the process of interest.

If we consider exclusive $\st_1$ production and all possible $\st_1$
decay chains the maximum asymmetry is diluted slightly to
$\sim12.5\%$. In the mSUGRA scenario considered in this paper one
should expect to see a $3\sigma$ effect at $\mathcal{L} = 500\
\mathrm{fb}^{-1}$ for phases in the range $0.5\,\pi \lesssim
\phi_{A_t}(\mathrm{mod}\; \pi) \lesssim 0.9\,\pi$. If general SUSY production is
considered, significant backgrounds to our signal process are present
and extra kinematical cuts are required to remove this
background. Even after these cuts some SUSY background remains and our
maximum asymmetry is reduced to $\sim6.5\%$. To see a $3\sigma$ effect
at the LHC would require $\mathcal{L} = 1\ \mathrm{ab}^{-1}$ of data
for sensitivity to phases in the range $0.6\,\pi \lesssim \phi_{A_t}(\mathrm{mod}\; \pi)
\lesssim 0.85\,\pi$.

We emphasise that the asymmetry after momentum reconstruction is a
much cleaner observable from a theoretical point of view, thanks to a
well defined final state. Therefore, using the above technique provides
 prospects for the observation of CP-violating effects for a range
of the phase $\phi_{A_t}$ after a few years of LHC running at the high
luminosity. The full assessment of LHC's ability to resolve CP
violation in the MSSM, however, will definitely require a detailed simulation
of detector effects, SM and SUSY backgrounds which is beyond the scope of the present
phenomenological analysis. The promising results of this study 
may  encourage such further simulations.




\section*{Acknowledgements}
The authors wish to thank Peter Wienemann, Philip Bechtle, Bj\"{o}rn Gosdzik and Frank Krauss for valuable discussions. We also are grateful to David Grellscheid and Peter Richardson for their assistance in the use of \texttt{Herwig++}. In addition we would like to thank Frank Siegert and Hendrik Hoeth for their help in the use of \texttt{Rivet}.
KR was supported by the EU Network MRTN-CT-2006-035505 ``Tools and Precision Calculations for Physics Discoveries at Colliders'' (HEPTools). JT was supported by the UK Science and Technology Facilities Council (STFC). The authors would like to thank the
Helmholtz Alliance HA-101 ``Physics at the Terascale'' for support.

\section*{Appendices}

\begin{appendix}

\section{Mixing in the stop sector}
\label{sec:StopSector}

In the Minimal Supersymmetric Standard Model the stop sector is defined by the mass matrix ${\cal M}_{\tilde{t}}$ in the basis of gauge eigenstates $(\tilde{t}_L, \tilde{t}_R)$. The $2 \times 2$ mass matrix depends on the soft scalar masses $M_{\tilde{Q}}$ and $M_{\tilde{U}}$, the supersymmetric higgsino mass parameter $\mu$, and the soft SUSY-breaking trilinear coupling $A_t$. It is given as \cite{Ellis:1983ed},
\begin{eqnarray} \label{eq:StopMix}
    {\cal M}^2_{\tilde{t}}=
     \left( \begin{array}{cc}
                m_t^2 + m_{LL}^2 & {m_{LR}^*}\, m_t \\
                {m_{LR}}\, m_t     & m_t^2 + m_{RR}^2
            \end{array} \right)\, ,
\end{eqnarray}
where,
\begin{eqnarray}
  m_{LL}^2 &=& M_{\tilde{Q}}^2
    + m_Z^2\cos 2\beta\,( \frac{1}{2} - \frac{2}{3} \sin^2\theta_W )\, , \\
  m_{RR}^2 & =&  M_{\tilde{U}}^2
                  + \frac{2}{3} m_Z^2 \cos 2\beta\, \sin^2\theta_W \, , \\
  m_{LR}    &= & {A_t} - {\mu^*} \cot\beta \,  ,
\end{eqnarray}
and $\tan\beta=v_2/v_1$ is the ratio of the vacuum expectation values of the
two neutral Higgs fields which break the electroweak symmetry. From the above parameters only $\mu$ and $A_t$ can take complex values,
\begin{eqnarray}\label{eq:at-mu}
A_t = |A_t|\, {\rm e}^{{\rm i}\phi_{A_t}},\qquad
\mu = |\mu|\, {\rm e}^{{\rm i}\phi_\mu}, \qquad (0\leq \phi_{A_t}, \phi_\mu < 2\pi)\,,
\end{eqnarray}
thus yielding CP violation in the stop sector. 

The hermitian matrix ${\cal M}^2_{\tilde{t}}$ is diagonalized by a unitary matrix $\mathcal{R}_{\tilde{t}}$,
\begin{eqnarray}
\mathcal{R}_{\tilde{t}}\: {\cal M}^2_{\tilde{t}}\: \mathcal{R}_{\tilde{t}}^\dag =
\left( \begin{array}{c c} m^2_{\tilde{t}_1} & 0 \\ 0 & m^2_{\tilde{t}_2} \end{array}
 \right) \, ,
\end{eqnarray}
where we choose the convention $m^2_{\tilde{t}_1} < m^2_{\tilde{t}_2} $ for the masses of $\tilde{t}_1$ and $\tilde{t}_2$.
The matrix $\mathcal{R}_{\tilde{t}}$ rotates the gauge eigenstates, $\tilde{t}_L$ and $\tilde{t}_R$, into the mass eigenstates $\tilde{t}_1$ and $\tilde{t}_2$ as follows,
\begin{eqnarray} \label{eq:stop-mix}
\left( \begin{array}{c}
                {\tilde{t}}_1 \\
                {\tilde{t}}_2
\end{array} \right) = \mathcal{R}_{\tilde{t}}
\left( \begin{array}{c}
{\tilde{t}}_L \\
{\tilde{t}}_R
\end{array} \right) =
\left( \begin{array}{cc}
 \cos\theta_{\tilde{t}}  &
 \sin\theta_{\tilde{t}}\, \mathrm{e}^{-\mathrm{i}\phi_{\tilde{t}}}   \\
- \sin\theta_{\tilde{t}}\, \mathrm{e}^{\mathrm{i}\phi_{\tilde{t}}}&
 \cos\theta_{\tilde{t}}
            \end{array} \right)
\left( \begin{array}{c}
                {\tilde{t}}_L \\
                {\tilde{t}}_R
\end{array} \right) \, ,
\end{eqnarray}
where $\theta_{\tilde{t}}$ and $\phi_{\tilde{t}}$ are the mixing angle and the CP-violating phase of the stop sector, respectively. The masses are given by,
\begin{eqnarray}
m_{\tilde{t}_{1,2}} = \frac{1}{2} \left( 2 m_t^2 + m_{LL}^2 + m_{RR}^2 \mp \sqrt{(m_{LL}^2 - m_{RR}^2)^2 + 4 |m_{LR}|^2 m_t^2} \right) \,,
\end{eqnarray}
whereas for the mixing angle and the CP phase we have,
\begin{eqnarray}
\cos\theta_{\tilde{t}} &=& \frac{-m_t|m_{LR}|}{\sqrt{m_t^2 |m_{LR}|^2+(m_{\tilde{t}_1}^2-m_{LL}^2)^2}}\,,\\
\sin\theta_{\tilde{t}} &=& \frac{m_{LL}^2-m_{\tilde{t}_1}^2}{\sqrt{m_t^2 |m_{LR}|^2+(m_{\tilde{t}_1}^2-m_{LL}^2)^2}}\,,\\
\phi_{\tilde{t}} &=& \arg (A_t - \mu^* \cot\beta)\, \label{eq:phtau}.
\end{eqnarray}
By convention we take $0\leq \theta_{\tilde{t}}<\pi$ and $0\leq \phi_{\tilde{t}} < 2 \pi$. It must be noted that $\phi_{\tilde{t}}$ is an \textquoteleft effective' phase and does not directly correspond to the phase of any MSSM parameter. Instead, the phase will have contributions from both $\phi_{A_t}$ and $\phi_{\mu}$. However, in this study we set $\phi_{\mu}=0$ due to the EDM constraints.

If $m_{LL} < m_{RR}$ then $\cos^2\theta_{\tilde{t}} > \frac{1}{{2}}$ and $\tilde{t}_1$ has a predominantly left gauge character. On the other hand, if $m_{LL} > m_{RR}$ then $\cos^2\theta_{\tilde{t}} < \frac{1}{{2}}$ and $\tilde{t}_1$ has a predominantly right gauge character.

\section{Mixing in the neutralino sector}
\label{sec:NeutSector}

In the MSSM, the four neutralinos $\tilde{\chi}^0_i$ ($i=1,2,3,4$) are
mixtures of the neutral $U(1)$ and $SU(2)$ gauginos, $\tilde{B}$ and
$\tilde{W}^3$, and the  higgsinos, $\tilde{H}^0_1$ and
$\tilde{H}^0_2$. The neutralino mass matrix in the $(\tilde{B},
\tilde{W}^3, \tilde{H}^0_1, \tilde{H}^0_2)$ basis \cite{Haber:1984rc,Bartl:1986hp},
\begin{eqnarray}
{\cal M}_N =\left(\begin{array}{cccc}
          M_1  &  0  & -m_Z c_\beta s_W   &  m_Z s_\beta s_W  \\[1mm]
          0    & M_2 & m_Z c_\beta c_W    & -m_Z s_\beta c_W  \\[1mm]
       -m_Z c_\beta s_W &  m_Z c_\beta c_W &   0  & -\mu \\[1mm]
        m_Z s_\beta s_W & -m_Z s_\beta c_W & -\mu &  0
               \end{array}\right)
\label{eq:mass_matrix}
\end{eqnarray}
is built up by the fundamental SUSY parameters: the $U(1)$ and $SU(2)$
gaugino masses $M_1$ and $M_2$, the higgsino mass parameter $\mu$, and
$\tan\beta=v_2/v_1$ ($c_\beta = \cos\beta$, $s_W = \sin\theta_W$ etc.). In addition to the $\mu$ parameter, a non-trivial CP phase can also
be attributed to the $M_1$ parameter:
\begin{eqnarray}
M_1 = |M_1|\, {\rm e}^{{\rm i}\phi_1},\quad \ \ (0\leq \phi_1 < 2\pi)\,.
\end{eqnarray}
Since the complex matrix ${\cal M}_N$ is symmetric, 
one unitary matrix $N$ is
sufficient to rotate the gauge eigenstate basis $(\tilde{B}, \tilde{W}^3,
\tilde{H}^0_1, \tilde{H}^0_2)$ to the mass eigenstate basis of the Majorana
fields~$\tilde{\chi}^0_i$
\begin{eqnarray}
{\rm diag}(m_{\tilde{\chi}_1^0},m_{\tilde{\chi}_2^0},m_{\tilde{\chi}_3^0},m_{\tilde{\chi}_4^0}) = N^*{\cal M}_N N^\dagger\,,\qquad (m_{\tilde{\chi}_1^0}<m_{\tilde{\chi}_2^0}<m_{\tilde{\chi}_3^0}<m_{\tilde{\chi}_4^0})\,. \label{eq:Nmatrix}
\end{eqnarray}
The masses  $m_{\tilde{\chi}_i^0}$ ($i=1,2,3,4$) can
be chosen to be real
and positive by a suitable definition of the unitary matrix
$N$.

\section{Interaction Lagrangian and couplings}
\label{sec:lagrangian-couplings}

The interaction Lagrangian for the stop decay $(\st_i \to \neu{j} t)$ is,
\begin{equation}
\mathcal{L}_{t \tilde{t} \tilde{\chi}^0} = {\tilde{\chi}^0_j} \left(a_{ij} P_L + b_{ij} P_R \right) t  \: \tilde{t}^*_i + \mathrm{h.c.} \,,
\end{equation}
where $P_{L,R}=\frac{1}{2}(1\mp\gamma_5)$. The couplings are given by,
\begin{eqnarray}
a_{ij} &=& -\frac{e}{\sqrt{2}\: s_W c_W}\: \mathcal{R}^{\tilde{t}}_{i1} \left( \frac{1}{3} s_W N^*_{j1} + c_W N^*_{j2} \right)  - Y_t\: \mathcal{R}^{\tilde{t}}_{i2} N^*_{j4}\,, \label{eq:a_ij}\\
b_{ij} &=& \frac{2 \sqrt{2}\: e}{3 c_W}\: \mathcal{R}^{\tilde{t}}_{i2} N_{j1} - Y_t\: \mathcal{R}^{\tilde{t}}_{i1} N_{j4}\,, \label{eq:b_ij}
\end{eqnarray}
where $\mathcal{R}^{\tilde{t}}_{ij}$ are the entries of stop mixing matrix, Eq.~(\ref{eq:stop-mix}), and $N_{ij}$ are the entries of the neutralino mixing matrix, Eq.~(\ref{eq:Nmatrix}). The top Yukawa coupling is given by,
\begin{eqnarray}
Y_t = \frac{e\: m_t}{\sqrt{2}\: m_W s_W \sin\beta}\,.
\end{eqnarray}
The interaction Lagrangian for the neutralino decay $(\neu{j} \to \sl \ell)$ is,
\begin{equation}
  \mathcal{L}_{\ell\tilde{\ell}\tilde{\chi}^0_j} =
 {g} f_{Lj}^{\ell}\: \bar{\ell}\: P_R\: \tilde{\chi}^0_j\: \tilde{\ell}_L + {g} f_{Rj}^{\ell}\: \bar{\ell}\: P_L\: \tilde{\chi}^0_j\: \tilde{\ell}_R +
 \textrm{h.c.}\,
\end{equation}
where $g = e/\sin\theta_W$. The couplings are given by,
 \begin{eqnarray}
  f_{Lj}^{\ell} &=& \frac{1}{\sqrt{2}} \, (\tan\t_W N_{j1} +  {N_{j2}})\,,  \\
  f_{Rj}^{\ell} &=& -\sqrt{2}\, \tan\t_W  N_{j1}^* \,, \\
\end{eqnarray}

\section{Amplitude squared including full spin correlations} \label{sect:AmpSq}
\subsection{Neutralino production $\tilde{t}_1 \to \tilde{\chi}^0_j t$\label{sect:stopdecay}}

Here we give the analytic expression for the neutralino production density matrix \cite{Bartl:2004jr},
\begin{equation}
|M(\tilde{t}_1\to \tilde{\chi}^0_j t)|^2=P(\tilde{\chi}^0_j t)+\Sigma^a_P(\tilde{\chi}^0_j)+
\Sigma^b_P(t)+\Sigma^{ab}_P(\tilde{\chi}^0_j t)\,,
\label{eq_mq}
\end{equation}
whose spin-independent contribution reads
\begin{equation}
P(\tilde{\chi}^0_j t)=(|a_{1j}|^2+|b_{1j}|^2)(p_t
p_{\tilde{\chi}^0_j})-2 m_t m_{\tilde{\chi}^0_j}
Re(a_{1j}b^*_{1j})\,,
\end{equation}
where $p_t$ and $p_{\tilde{\chi}^0_k}$ denote the four-momenta of the $t$-quark and 
the neutralino $\tilde{\chi}^0_k$. The coupling constants $a_{ij}$ and $b_{ij}$ are shown in Eq.~(\ref{eq:a_ij},\ref{eq:b_ij}) and by substituting the explicit matrix elements of Eq.~(\ref{eq:stop-mix}) we can show the specific parameter dependence \cite{Rolbiecki:2009hk}, 
\begin{eqnarray}
\label{eq:neuq2} 
 |a_{1j}|^2 &+& |b_{1j}|^2  = \nonumber\\ &  = & \cos^2\theta_{\st} \left( \frac{e^2}{2 s_W^2 c_W^2 } \Big| \frac{1}{3} s_W N_{j1} + c_W N_{j2} \Big|^2 + Y_t^2 |N_{j4}|^2 \right)
+ \sin^2\theta_{\st} \left(\frac{8 e^2}{9 c_W^2} |N_{j1}|^2 + Y_t^2 |N_{j4}|^2 \right)\nonumber \\ 
&  + & 2 \sin\theta_{\st} \cos\theta_{\st}\: Y_t \left( \frac{e}{\sqrt{2}\: s_W c_W}\: {\rm Re} \bigg[ {\rm e}^{{\rm i} \phi_{\st}} \left( \frac{1}{3} s_W N^*_{j1} + c_W N^*_{j2} \right) N_{j4} \bigg] \right.
\left. -\frac{2\sqrt{2}\:e}{3 c_W}\: {\rm Re} \big[ {\rm e}^{-{\rm i} \phi_{\st}} N_{j1} N_{j4}^* \big] \right)\,.
\nonumber \\
\end{eqnarray}
\begin{eqnarray}
\label{eq:neuqlqr}
{\rm Re}\left[ a_{1j} b_{1j}^{*} \right]  & =&\cos^2\theta_{\st}\: \frac{e}{\sqrt{2}\: s_W c_W} Y_t\: {\rm Re} \left[\left( \frac{1}{3} s_W N^*_{j1} + c_W N^*_{j2} \right) N_{j4}^* \right] + \sin^2\theta_{\st}\: \frac{2\sqrt{2}\: e}{3 c_W}\: Y_t\: {\rm Re}[N_{j4}^* N_{j1}^*] \nonumber \\
& + & \sin\theta_{\st} \cos\theta_{\st} \left( Y_t^2\: {\rm Re} \left[{\rm e}^{-{\rm i} \phi_{\st}} N_{j4}^{*2} \right] - \frac{2}{3} \frac{e^2}{s_W c_W^2}\: {\rm Re} \left[  {\rm e}^{{\rm i} \phi_{\st}} \left( \frac{1}{3} s_W N^*_{j1} + c_W N^*_{j2} \right) N_{j1}^* \right] \right)\,. \nonumber \\
\end{eqnarray}

The spin-dependent terms that depend on individual spin contributions are T-even and are given by,
\begin{eqnarray}
\Sigma^a_P(\tilde{\chi}^0_j) &=& (|b_{ij}|^2-|a_{ij}|^2)m_{\tilde{\chi}^0_j} 
(p_t s^a(\tilde{\chi}^0_j))\,,\label{eq_prod-ea}\\
\Sigma^b_P(t) &=& (|b_{ij}|^2-|a_{ij}|^2)m_{t} 
(p_{\tilde{\chi}^0_j} s^b(t))\,,\label{eq_prod-eb}
\end{eqnarray}
where $s^a(\tilde{\chi}^0_j)$ $(s^b(t))$ denote the spin-basis vectors of the neutralino
$\tilde{\chi}^0_j$ (t-quark). Again the coupling constants can be expanded as,
\begin{eqnarray}
\label{eq:neua-b} 
 |b_{1j}|^2 &-& |a_{1j}|^2  = \nonumber\\ &  = & \cos^2\theta_{\st} \left(Y_t^2 |N_{j4}|^2 - \frac{e^2}{2 s_W^2 c_W^2 } \Big| \frac{1}{3} s_W N_{j1} + c_W N_{j2} \Big|^2 \right)
+ \sin^2\theta_{\st} \left(\frac{8 e^2}{9 c_W^2} |N_{j1}|^2 - Y_t^2 |N_{j4}|^2 \right)\nonumber \\ 
&  - & 2 \sin\theta_{\st} \cos\theta_{\st}\: Y_t \left( \frac{e}{\sqrt{2}\: s_W c_W}\: {\rm Re} \bigg[ {\rm e}^{{\rm i} \phi_{\st}} \left( \frac{1}{3} s_W N^*_{j1} + c_W N^*_{j2} \right) N_{j4} \bigg] \right.
\left. +\frac{2\sqrt{2}\:e}{3 c_W}\: {\rm Re} \big[ {\rm e}^{-{\rm i} \phi_{\st}} N_{j1} N_{j4}^* \big] \right)\,.
\nonumber \\
\end{eqnarray}
The terms that depend simultaneously on the spin of the top quark and of the neutralino can be split into
T-even, $\Sigma^{ab}_{P,even}(\tilde{\chi}^0_j t)$, and T-odd, 
$\Sigma^{ab}_{P,odd}(\tilde{\chi}^0_j t)$. The T-even contributions are as follows,
\begin{eqnarray}
\Sigma^{ab}_{P,even}(\tilde{\chi}^0_j t)&=& 2
Re(a_{ij}b^*_{ij}) [(s^a(\tilde{\chi}^0_j) p_t) (s^b(t)
p_{\tilde{\chi}^0_j}) - (p_t p_{\tilde{\chi}^0_j})
(s^a(\tilde{\chi}^0_j) s^b(t))]\nonumber\\ 
& + & m_t m_{\tilde{\chi}^0_j} (s^a(\tilde{\chi}^0_j)s^b(t))
(|a_{ij}|^2+|b_{ij}|^2)\,. \label{eq_prod-e}
\end{eqnarray}
The T-odd contributions that generate the triple product correlations that we are interested in are,
\begin{equation}
 \Sigma^{ab}_{P,odd}(\tilde{\chi}^0_j t)= - g^2
Im(a_{ij}b^*_{ij}) f_4^{ab}\,, \label{eq_prod-o}
\end{equation}
where the T-odd kinematical factor is given by,
\begin{equation}
f_4^{ab}= \epsilon_{\mu\nu\rho\sigma}s^{a,\mu}(\tilde{\chi}^0_j)p^{\nu}_{\tilde{\chi}^0_j}
s^{b,\rho}(t)p^{\sigma}_t\,.
\label{eq_f4}
\end{equation}
Sec.\ref{sec:Todd_Struc} explains how this epsilon product generates the triple product observable. We again expand the coupling constant to see the functional dependence,
\begin{eqnarray}
\label{eq:ImCoup}
{\rm Im}\left[ a_{1j} b_{1j}^{*} \right]  & =&\cos^2\theta_{\st}\: \frac{e}{\sqrt{2}\: s_W c_W} Y_t\: {\rm Im} \left[\left( \frac{1}{3} s_W N^*_{j1} + c_W N^*_{j2} \right) N_{j4}^* \right] \nonumber \\
& + & \sin^2\theta_{\st}\: \frac{2\sqrt{2}\: e}{3 c_W}\: Y_t\: {\rm Im}[N_{j4}^* N_{j1}^*] \nonumber \\
& + & \sin\theta_{\st} \cos\theta_{\st} \Bigg{(} Y_t^2\: {\rm Im} \left[{\rm e}^{-{\rm i} \phi_{\st}} N_{j4}^{*2} \right] \nonumber \\
& \;\; & \;\;\;\;\;\;\;\;\;\;\;\;\;\;\;\;\;\;\;\; - \;\;  \frac{2}{3} \frac{e^2}{s_W c_W^2}\: {\rm Im} \left[  {\rm e}^{{\rm i} \phi_{\st}} \left( \frac{1}{3} s_W N^*_{j1} + c_W N^*_{j2} \right) N_{j1}^* \right] \Bigg)\,. \nonumber \\
\end{eqnarray}

\subsection{Neutralino decay  $\neu{2} \to \sl^+_R \ell^-$ \label{sect:neut2decay}}
We provide analytical expressions for the 2-body decay of the $\neu{2}$
into a $\sl^+_R$ and the final-state $\ell^-$ \cite{Choi:1999cc},
\begin{equation}
D(\neu{2})=\frac{g^2}{4} |f^l_{L2}|^2 \{ m_{\neu{2}}^2 - m_{\tilde{\ell}_R}^2 \}.
\end{equation}
The spin-dependent contribution is T-even and reads:
\begin{equation}
\Sigma_D^{a}(\neu{2})= \frac{g^2}{2} |f^l_{L2}|^2 m_{\neu{2}} \{ (s^a(\neu{2}) p_{\ell^-} \}.
\label{eq_neutdecay}
\end{equation}

\subsection{Top decay  $t \to W^+ b$ \label{sect:topdecay}}

We provide analytical expressions for the 2-body decay of the top quark 
into a $W$-boson and the final-state bottom quark \cite{Fischer:2001gp},
\begin{equation}
D(t)=\frac{g^2}{4} \{ m_t^2 -2 m_W^2+\frac{m_t^4}{m_W^2} \}.
\end{equation}
The spin-dependent contribution is T-even and reads:
\begin{equation}
\Sigma_D^{b}(t)=-\frac{g^2}{2} m_t \{ (s^b(t) p_b) + \frac{m_t^2 -m_W^2}{m_W^2}
(s^b(t) p_W) \}.
\label{eq_stdecay}
\end{equation}

\end{appendix}


\addcontentsline{toc}{section}{References}
\bibliography{refs}

\providecommand{\href}[2]{#2}\begingroup\raggedright\begin{thebibliography}{10}

\bibitem{Buchmueller:2009fn}
O.~Buchmueller {\em et al.}, ``{Likelihood Functions for Supersymmetric
  Observables in Frequentist Analyses of the CMSSM and NUHM1},''
\href{http://arxiv.org/abs/0907.5568}{{\tt arXiv:0907.5568 [hep-ph]}}.

\bibitem{Cohen:1993nk}
A.~G. Cohen, D.~B. Kaplan, and A.~E. Nelson, ``{Progress in electroweak
  baryogenesis},'' {\em Ann. Rev. Nucl. Part. Sci.} {\bf 43} (1993)  27--70,
\href{http://arxiv.org/abs/hep-ph/9302210}{{\tt arXiv:hep-ph/9302210}}.

\bibitem{Gavela:1994dt}
M.~B. Gavela, P.~Hernandez, J.~Orloff, O.~Pene, and C.~Quimbay, ``{Standard
  model CP violation and baryon asymmetry. Part 2: Finite temperature},''
  \href{http://dx.doi.org/10.1016/0550-3213(94)00410-2}{{\em Nucl. Phys.} {\bf
  B430} (1994)  382--426},
\href{http://arxiv.org/abs/hep-ph/9406289}{{\tt arXiv:hep-ph/9406289}}.

\bibitem{Rubakov:1996vz}
V.~A. Rubakov and M.~E. Shaposhnikov, ``{Electroweak baryon number
  non-conservation in the early universe and in high-energy collisions},'' {\em
  Usp. Fiz. Nauk} {\bf 166} (1996)  493--537,
\href{http://arxiv.org/abs/hep-ph/9603208}{{\tt arXiv:hep-ph/9603208}}.

\bibitem{Dimopoulos:1995ju}
S.~Dimopoulos and D.~W. Sutter, ``{The Supersymmetric flavor problem},''
  \href{http://dx.doi.org/10.1016/0550-3213(95)00421-N}{{\em Nucl. Phys.} {\bf
  B452} (1995)  496--512},
\href{http://arxiv.org/abs/hep-ph/9504415}{{\tt arXiv:hep-ph/9504415}}.

\bibitem{Ibrahim:2007fb}
T.~Ibrahim and P.~Nath, ``{CP violation from standard model to strings},''
  \href{http://dx.doi.org/10.1103//RevModPhys.80.577}{{\em Rev. Mod. Phys.}
  {\bf 80} (2008)  577--631},
\href{http://arxiv.org/abs/0705.2008}{{\tt arXiv:0705.2008 [hep-ph]}}.

\bibitem{Ellis:2008zy}
J.~R. Ellis, J.~S. Lee, and A.~Pilaftsis, ``{Electric Dipole Moments in the
  MSSM Reloaded},'' \href{http://dx.doi.org/10.1088/1126-6708/2008/10/049}{{\em
  JHEP} {\bf 10} (2008)  049},
\href{http://arxiv.org/abs/0808.1819}{{\tt arXiv:0808.1819 [hep-ph]}}.

\bibitem{Kizukuri:1992nj}
Y.~Kizukuri and N.~Oshimo, ``{The Neutron and electron electric dipole moments
  in supersymmetric theories},''
\href{http://dx.doi.org/10.1103/PhysRevD.46.3025}{{\em Phys. Rev.} {\bf D46}
  (1992)  3025--3033}.

\bibitem{Ibrahim:1998je}
T.~Ibrahim and P.~Nath, ``{The neutron and the lepton EDMs in MSSM, large CP
  violating phases, and the cancellation mechanism},''
  \href{http://dx.doi.org/10.1103/PhysRevD.58.111301}{{\em Phys. Rev.} {\bf
  D58} (1998)  111301},
\href{http://arxiv.org/abs/hep-ph/9807501}{{\tt arXiv:hep-ph/9807501}}.

\bibitem{Ibrahim:1999af}
T.~Ibrahim and P.~Nath, ``{Large CP phases and the cancellation mechanism in
  EDMs in SUSY, string and brane models},''
  \href{http://dx.doi.org/10.1103/PhysRevD.61.093004}{{\em Phys. Rev.} {\bf
  D61} (2000)  093004},
\href{http://arxiv.org/abs/hep-ph/9910553}{{\tt arXiv:hep-ph/9910553}}.

\bibitem{Brhlik:1998zn}
M.~Brhlik, G.~J. Good, and G.~L. Kane, ``{Electric dipole moments do not
  require the CP-violating phases of supersymmetry to be small},''
  \href{http://dx.doi.org/10.1103/PhysRevD.59.115004}{{\em Phys. Rev.} {\bf
  D59} (1999)  115004},
\href{http://arxiv.org/abs/hep-ph/9810457}{{\tt arXiv:hep-ph/9810457}}.

\bibitem{Abel:2001vy}
S.~Abel, S.~Khalil, and O.~Lebedev, ``{EDM constraints in supersymmetric
  theories},'' \href{http://dx.doi.org/10.1016/S0550-3213(01)00233-4}{{\em
  Nucl. Phys.} {\bf B606} (2001)  151--182},
\href{http://arxiv.org/abs/hep-ph/0103320}{{\tt arXiv:hep-ph/0103320}}.

\bibitem{Arnowitt:2001pm}
R.~L. Arnowitt, B.~Dutta, and Y.~Santoso, ``{SUSY phases, the electron electric
  dipole moment and the muon magnetic moment},''
  \href{http://dx.doi.org/10.1103/PhysRevD.64.113010}{{\em Phys. Rev.} {\bf
  D64} (2001)  113010},
\href{http://arxiv.org/abs/hep-ph/0106089}{{\tt arXiv:hep-ph/0106089}}.

\bibitem{Li:2010ax}
Y.~Li, S.~Profumo, and M.~Ramsey-Musolf, ``{A Comprehensive Analysis of
  Electric Dipole Moment Constraints on CP-violating Phases in the MSSM},''
\href{http://arxiv.org/abs/1006.1440}{{\tt arXiv:1006.1440 [hep-ph]}}.

\bibitem{Lee:2003nta}
J.~S. Lee {\em et al.}, ``{CPsuperH: A computational tool for Higgs
  phenomenology in the minimal supersymmetric standard model with explicit CP
  violation},'' \href{http://dx.doi.org/10.1016/S0010-4655(03)00463-6}{{\em
  Comput. Phys. Commun.} {\bf 156} (2004)  283--317},
\href{http://arxiv.org/abs/hep-ph/0307377}{{\tt arXiv:hep-ph/0307377}}.

\bibitem{Lee:2007gn}
J.~S. Lee, M.~Carena, J.~Ellis, A.~Pilaftsis, and C.~E.~M. Wagner,
  ``{CPsuperH2.0: an Improved Computational Tool for Higgs Phenomenology in the
  MSSM with Explicit CP Violation},''
  \href{http://dx.doi.org/10.1016/j.cpc.2008.09.003}{{\em Comput. Phys.
  Commun.} {\bf 180} (2009)  312--331},
\href{http://arxiv.org/abs/0712.2360}{{\tt arXiv:0712.2360 [hep-ph]}}.

\bibitem{Deppisch:2009nj}
F.~Deppisch and O.~Kittel, ``{Probing SUSY CP Violation in Two-Body Stop Decays
  at the LHC},''
\href{http://arxiv.org/abs/0905.3088}{{\tt arXiv:0905.3088 [hep-ph]}}.

\bibitem{Kraml:2007pr}
S.~Kraml, ``{CP violation in SUSY},''
\href{http://arxiv.org/abs/0710.5117}{{\tt arXiv:0710.5117 [hep-ph]}}.

\bibitem{Barger:1999tn}
V.~D. Barger, T.~Han, T.-J. Li, and T.~Plehn, ``{Measuring CP Violating Phases
  at a Future Linear Collider},''
  \href{http://dx.doi.org/10.1016/S0370-2693(00)00070-8}{{\em Phys. Lett.} {\bf
  B475} (2000)  342--350},
\href{http://arxiv.org/abs/hep-ph/9907425}{{\tt arXiv:hep-ph/9907425}}.

\bibitem{Kneur:1999nx}
J.~L. Kneur and G.~Moultaka, ``{Phases in the gaugino sector: Direct
  reconstruction of the basic parameters and impact on the neutralino pair
  production},'' \href{http://dx.doi.org/10.1103/PhysRevD.61.095003}{{\em Phys.
  Rev.} {\bf D61} (2000)  095003},
\href{http://arxiv.org/abs/hep-ph/9907360}{{\tt arXiv:hep-ph/9907360}}.

\bibitem{Kittel:2009fg}
O.~Kittel, ``{SUSY CP phases and asymmetries at colliders},''
\href{http://arxiv.org/abs/0904.3241}{{\tt arXiv:0904.3241 [hep-ph]}}.

\bibitem{Hesselbach:2007dq}
S.~Hesselbach, ``{CP Violation in SUSY Particle Production and Decay},''
\href{http://arxiv.org/abs/0709.2679}{{\tt arXiv:0709.2679 [hep-ph]}}.

\bibitem{Bartl:2004jr}
A.~Bartl, E.~Christova, K.~Hohenwarter-Sodek, and T.~Kernreiter, ``{Triple
  product correlations in top squark decays},''
  \href{http://dx.doi.org/10.1103/PhysRevD.70.095007}{{\em Phys. Rev.} {\bf
  D70} (2004)  095007},
\href{http://arxiv.org/abs/hep-ph/0409060}{{\tt arXiv:hep-ph/0409060}}.

\bibitem{Langacker:2007ur}
P.~Langacker, G.~Paz, L.-T. Wang, and I.~Yavin, ``{A T-odd observable sensitive
  to CP violating phases in squark decay},''
  \href{http://dx.doi.org/10.1088/1126-6708/2007/07/055}{{\em JHEP} {\bf 07}
  (2007)  055},
\href{http://arxiv.org/abs/hep-ph/0702068}{{\tt arXiv:hep-ph/0702068}}.

\bibitem{Ellis:2008hq}
J.~Ellis, F.~Moortgat, G.~Moortgat-Pick, J.~M. Smillie, and J.~Tattersall,
  ``{Measurement of CP Violation in Stop Cascade Decays at the LHC},''
  \href{http://dx.doi.org/10.1140/epjc/s10052-009-0964-8}{{\em Eur. Phys. J.}
  {\bf C60} (2009)  633--651},
\href{http://arxiv.org/abs/0809.1607}{{\tt arXiv:0809.1607 [hep-ph]}}.

\bibitem{Kiers:2006aq}
K.~Kiers, A.~Szynkman, and D.~London, ``{CP violation in supersymmetric
  theories: stop(2) $\rightarrow$ stop(1) tau- tau+},''
  \href{http://dx.doi.org/10.1103/PhysRevD.74.035004}{{\em Phys. Rev.} {\bf
  D74} (2006)  035004},
\href{http://arxiv.org/abs/hep-ph/0605123}{{\tt arXiv:hep-ph/0605123}}.

\bibitem{Bartl:2006hh}
A.~Bartl, E.~Christova, K.~Hohenwarter-Sodek, and T.~Kernreiter, ``{CP
  asymmetries in scalar bottom quark decays},'' {\em JHEP} {\bf 11} (2006)
  076,
\href{http://arxiv.org/abs/hep-ph/0610234}{{\tt arXiv:hep-ph/0610234}}.

\bibitem{Deppisch:2010nc}
F.~F. Deppisch and O.~Kittel, ``{CP violation in sbottom decays},''
  \href{http://dx.doi.org/10.1007/JHEP06(2010)067}{{\em JHEP} {\bf 06} (2010)
  067},
\href{http://arxiv.org/abs/1003.5186}{{\tt arXiv:1003.5186 [hep-ph]}}.

\bibitem{MoortgatPick:2009jy}
G.~Moortgat-Pick, K.~Rolbiecki, J.~Tattersall, and P.~Wienemann, ``{Probing CP
  Violation with and without Momentum Reconstruction at the LHC},''
  \href{http://dx.doi.org/10.1007/JHEP01(2010)004}{{\em JHEP} {\bf 01} (2010)
  004},
\href{http://arxiv.org/abs/0908.2631}{{\tt arXiv:0908.2631 [hep-ph]}}.

\bibitem{Boos:2003vf}
E.~Boos {\em et al.}, ``{Polarisation in sfermion decays: Determining tan(beta)
  and trilinear couplings},''
  \href{http://dx.doi.org/10.1140/epjc/s2003-01288-y}{{\em Eur. Phys. J.} {\bf
  C30} (2003)  395--407},
\href{http://arxiv.org/abs/hep-ph/0303110}{{\tt arXiv:hep-ph/0303110}}.

\bibitem{Weiglein:2004hn}
{\bf LHC/LC Study Group} Collaboration, G.~Weiglein {\em et al.}, ``{Physics
  interplay of the LHC and the ILC},''
  \href{http://dx.doi.org/10.1016/j.physrep.2005.12.003}{{\em Phys. Rept.} {\bf
  426} (2006)  47--358},
\href{http://arxiv.org/abs/hep-ph/0410364}{{\tt arXiv:hep-ph/0410364}}.

\bibitem{Dreiner:2010ib}
H.~K. Dreiner, O.~Kittel, and A.~Marold, ``{Normal tau polarisation as a
  sensitive probe of CP violation in chargino decay},''
  \href{http://dx.doi.org/10.1103/PhysRevD.82.116005}{{\em Phys. Rev.} {\bf
  D82} (2010)  116005},
\href{http://arxiv.org/abs/1001.4714}{{\tt arXiv:1001.4714 [hep-ph]}}.

\bibitem{Dreiner:2010wj}
H.~Dreiner, O.~Kittel, S.~Kulkarni, and A.~Marold, ``{Testing the CP-violating
  MSSM in stau decays at the LHC and ILC},''
\href{http://arxiv.org/abs/1011.2449}{{\tt arXiv:1011.2449 [hep-ph]}}.

\bibitem{Nattermann:2009gh}
T.~Nattermann, K.~Desch, P.~Wienemann, and C.~Zendler, ``{Measuring
  tau-polarisation in Neutralino2 decays at the LHC},''
  \href{http://dx.doi.org/10.1088/1126-6708/2009/04/057}{{\em JHEP} {\bf 04}
  (2009)  057},
\href{http://arxiv.org/abs/0903.0714}{{\tt arXiv:0903.0714 [hep-ph]}}.

\bibitem{Atwood:2000tu}
D.~Atwood, S.~Bar-Shalom, G.~Eilam, and A.~Soni, ``{CP violation in top
  physics},'' \href{http://dx.doi.org/10.1016/S0370-1573(00)00112-5}{{\em Phys.
  Rept.} {\bf 347} (2001)  1--222},
\href{http://arxiv.org/abs/hep-ph/0006032}{{\tt arXiv:hep-ph/0006032}}.

\bibitem{Abazov:2010hv}
{\bf D0} Collaboration, V.~M. Abazov {\em et al.}, ``{Evidence for an anomalous
  like-sign dimuon charge asymmetry},''
  \href{http://dx.doi.org/10.1103/PhysRevD.82.032001}{{\em Phys. Rev.} {\bf
  D82} (2010)  032001},
\href{http://arxiv.org/abs/1005.2757}{{\tt arXiv:1005.2757 [hep-ex]}}.

\bibitem{Aaltonen:2011kc}
{\bf The CDF} Collaboration, T.~Aaltonen {\em et al.}, ``{Evidence for a Mass
  Dependent Forward-Backward Asymmetry in Top Quark Pair Production},''
\href{http://arxiv.org/abs/1101.0034}{{\tt arXiv:1101.0034 [hep-ex]}}.

\bibitem{MoortgatPick:1999di}
G.~A. Moortgat-Pick, H.~Fraas, A.~Bartl, and W.~Majerotto, ``{Polarization and
  spin effects in neutralino production and decay},''
  \href{http://dx.doi.org/10.1007/s100529900076}{{\em Eur. Phys. J.} {\bf C9}
  (1999)  521--534},
\href{http://arxiv.org/abs/hep-ph/9903220}{{\tt arXiv:hep-ph/9903220}}.

\bibitem{Haber:1994pe}
H.~E. Haber, ``{Spin formalism and applications to new physics searches},''
\href{http://arxiv.org/abs/hep-ph/9405376}{{\tt arXiv:hep-ph/9405376}}.

\bibitem{Schwinger:1951xk}
J.~S. Schwinger, ``{The theory of quantized fields. I},''
\href{http://dx.doi.org/10.1103/PhysRev.82.914}{{\em Phys. Rev.} {\bf 82}
  (1951)  914--927}.

\bibitem{Schwinger:1953tb}
J.~S. Schwinger, ``{The theory of quantized fields. II},''
\href{http://dx.doi.org/10.1103/PhysRev.91.713}{{\em Phys. Rev.} {\bf 91}
  (1953)  713--728}.

\bibitem{Kawagoe:2004rz}
K.~Kawagoe, M.~M. Nojiri, and G.~Polesello, ``{A new SUSY mass reconstruction
  method at the CERN LHC},''
  \href{http://dx.doi.org/10.1103/PhysRevD.71.035008}{{\em Phys. Rev.} {\bf
  D71} (2005)  035008},
\href{http://arxiv.org/abs/hep-ph/0410160}{{\tt arXiv:hep-ph/0410160}}.

\bibitem{Gjelsten:2004ki}
B.~K. Gjelsten, D.~J. Miller, 2, and P.~Osland, ``{Measurement of SUSY masses
  via cascade decays for SPS 1a},''
  \href{http://dx.doi.org/10.1088/1126-6708/2004/12/003}{{\em JHEP} {\bf 12}
  (2004)  003},
\href{http://arxiv.org/abs/hep-ph/0410303}{{\tt arXiv:hep-ph/0410303}}.

\bibitem{Nojiri:2003tu}
M.~M. Nojiri, G.~Polesello, and D.~R. Tovey, ``{Proposal for a new
  reconstruction technique for SUSY processes at the LHC},''
\href{http://arxiv.org/abs/hep-ph/0312317}{{\tt arXiv:hep-ph/0312317}}.

\bibitem{Nojiri:2007pq}
M.~M. Nojiri, G.~Polesello, and D.~R. Tovey, ``{A hybrid method for determining
  SUSY particle masses at the LHC with fully identified cascade decays},''
  \href{http://dx.doi.org/10.1088/1126-6708/2008/05/014}{{\em JHEP} {\bf 05}
  (2008)  014},
\href{http://arxiv.org/abs/0712.2718}{{\tt arXiv:0712.2718 [hep-ph]}}.

\bibitem{Cheng:2007xv}
H.-C. Cheng, J.~F. Gunion, Z.~Han, G.~Marandella, and B.~McElrath, ``{Mass
  Determination in SUSY-like Events with Missing Energy},''
  \href{http://dx.doi.org/10.1088/1126-6708/2007/12/076}{{\em JHEP} {\bf 12}
  (2007)  076},
\href{http://arxiv.org/abs/0707.0030}{{\tt arXiv:0707.0030 [hep-ph]}}.

\bibitem{Cheng:2009fw}
H.-C. Cheng, J.~F. Gunion, Z.~Han, and B.~McElrath, ``{Accurate Mass
  Determinations in Decay Chains with Missing Energy: II},''
\href{http://arxiv.org/abs/0905.1344}{{\tt arXiv:0905.1344 [hep-ph]}}.

\bibitem{Casadei:2010nf}
D.~Casadei, R.~Djilkibaev, and R.~Konoplich, ``{Reconstruction of stop quark
  mass at the LHC},''
\href{http://arxiv.org/abs/1006.5875}{{\tt arXiv:1006.5875 [hep-ph]}}.

\bibitem{Barr:2010zj}
A.~J. Barr and C.~G. Lester, ``{A Review of the Mass Measurement Techniques
  proposed for the Large Hadron Collider},''
\href{http://arxiv.org/abs/1004.2732}{{\tt arXiv:1004.2732 [hep-ph]}}.

\bibitem{Brooijmans:2010tn}
G.~Brooijmans {\em et al.}, ``{New Physics at the LHC. A Les Houches Report:
  Physics at TeV Colliders 2009 - New Physics Working Group},''
\href{http://arxiv.org/abs/1005.1229}{{\tt arXiv:1005.1229 [hep-ph]}}.

\bibitem{Porod:2003um}
W.~Porod, ``{SPheno, a program for calculating supersymmetric spectra, SUSY
  particle decays and SUSY particle production at e+ e- colliders},''
  \href{http://dx.doi.org/10.1016/S0010-4655(03)00222-4}{{\em Comput. Phys.
  Commun.} {\bf 153} (2003)  275--315},
\href{http://arxiv.org/abs/hep-ph/0301101}{{\tt arXiv:hep-ph/0301101}}.

\bibitem{Khachatryan:2011tk}
{\bf CMS} Collaboration, V.~Khachatryan {\em et al.}, ``{Search for
  Supersymmetry in pp Collisions at 7 TeV in Events with Jets and Missing
  Transverse Energy},''
  \href{http://dx.doi.org/10.1016/j.physletb.2011.03.021}{{\em Phys. Lett.}
  {\bf B698} (2011)  196--218},
\href{http://arxiv.org/abs/1101.1628}{{\tt arXiv:1101.1628 [hep-ex]}}.

\bibitem{daCosta:2011qk}
{\bf Atlas} Collaboration, J.~B.~G. da~Costa {\em et al.}, ``{Search for
  squarks and gluinos using final states with jets and missing transverse
  momentum with the ATLAS detector in sqrt(s) = 7 TeV proton-proton
  collisions},''
\href{http://arxiv.org/abs/1102.5290}{{\tt arXiv:1102.5290 [hep-ex]}}.

\bibitem{Nakamura:2010zzi}
{\bf Particle Data Group} Collaboration, K.~Nakamura {\em et al.}, ``{Review of
  particle physics},''
\href{http://dx.doi.org/10.1088/0954-3899/37/7A/075021}{{\em J. Phys.} {\bf
  G37} (2010)  075021}.

\bibitem{Bahr:2008pv}
M.~Bahr {\em et al.}, ``{Herwig++ Physics and Manual},''
\href{http://arxiv.org/abs/0803.0883}{{\tt arXiv:0803.0883 [hep-ph]}}.

\bibitem{Bahr:2008tf}
M.~Bahr {\em et al.}, ``{Herwig++ 2.3 Release Note},''
\href{http://arxiv.org/abs/0812.0529}{{\tt arXiv:0812.0529 [hep-ph]}}.

\bibitem{Martin:2007bv}
A.~D. Martin, W.~J. Stirling, R.~S. Thorne, and G.~Watt, ``{Update of Parton
  Distributions at NNLO},''
  \href{http://dx.doi.org/10.1016/j.physletb.2007.07.040}{{\em Phys. Lett.}
  {\bf B652} (2007)  292--299},
\href{http://arxiv.org/abs/0706.0459}{{\tt arXiv:0706.0459 [hep-ph]}}.

\bibitem{Richardson:2001df}
P.~Richardson, ``{Spin correlations in Monte Carlo simulations},'' {\em JHEP}
  {\bf 11} (2001)  029,
\href{http://arxiv.org/abs/hep-ph/0110108}{{\tt arXiv:hep-ph/0110108}}.

\bibitem{Buckley:2010ar}
A.~Buckley {\em et al.}, ``{Rivet user manual},''
\href{http://arxiv.org/abs/1003.0694}{{\tt arXiv:1003.0694 [hep-ph]}}.

\bibitem{Waugh:2006ip}
B.~M. Waugh {\em et al.}, ``{HZTool and Rivet: Toolkit and framework for the
  comparison of simulated final states and data at colliders},''
\href{http://arxiv.org/abs/hep-ph/0605034}{{\tt arXiv:hep-ph/0605034}}.

\bibitem{Cacciari:2005hq}
M.~Cacciari and G.~P. Salam, ``{Dispelling the $N^{3}$ myth for the $k_t$
  jet-finder},'' \href{http://dx.doi.org/10.1016/j.physletb.2006.08.037}{{\em
  Phys. Lett.} {\bf B641} (2006)  57--61},
\href{http://arxiv.org/abs/hep-ph/0512210}{{\tt arXiv:hep-ph/0512210}}.

\bibitem{Cacciari:2008gp}
M.~Cacciari, G.~P. Salam, and G.~Soyez, ``{The anti-$\mathrm{k_t}$ jet
  clustering algorithm},''
  \href{http://dx.doi.org/10.1088/1126-6708/2008/04/063}{{\em JHEP} {\bf 04}
  (2008)  063},
\href{http://arxiv.org/abs/0802.1189}{{\tt arXiv:0802.1189 [hep-ph]}}.

\bibitem{Aad:2009wy}
{\bf The ATLAS} Collaboration, G.~Aad {\em et al.}, ``{Expected Performance of
  the ATLAS Experiment - Detector, Trigger and Physics},''
\href{http://arxiv.org/abs/0901.0512}{{\tt arXiv:0901.0512 [hep-ex]}}.

\bibitem{Alwall:2007st}
J.~Alwall {\em et al.}, ``{MadGraph/MadEvent v4: The New Web Generation},''
  {\em JHEP} {\bf 09} (2007)  028,
\href{http://arxiv.org/abs/0706.2334}{{\tt arXiv:0706.2334 [hep-ph]}}.

\bibitem{Desch:2006xp}
K.~Desch, J.~Kalinowski, G.~Moortgat-Pick, K.~Rolbiecki, and W.~J. Stirling,
  ``{Combined LHC / ILC analysis of a SUSY scenario with heavy sfermions},''
  {\em JHEP} {\bf 12} (2006)  007,
\href{http://arxiv.org/abs/hep-ph/0607104}{{\tt arXiv:hep-ph/0607104}}.

\bibitem{Beenakker:1996ed}
W.~Beenakker, R.~Hopker, and M.~Spira, ``{PROSPINO: A program for the
  PROduction of Supersymmetric Particles In Next-to-leading Order QCD},''
\href{http://arxiv.org/abs/hep-ph/9611232}{{\tt arXiv:hep-ph/9611232}}.

\bibitem{Beenakker:1996ch}
W.~Beenakker, R.~Hopker, M.~Spira, and P.~M. Zerwas, ``{Squark and gluino
  production at hadron colliders},''
  \href{http://dx.doi.org/10.1016/S0550-3213(97)00084-9}{{\em Nucl. Phys.} {\bf
  B492} (1997)  51--103},
\href{http://arxiv.org/abs/hep-ph/9610490}{{\tt arXiv:hep-ph/9610490}}.

\bibitem{Beenakker:1997ut}
W.~Beenakker, M.~Kramer, T.~Plehn, M.~Spira, and P.~M. Zerwas, ``{Stop
  production at hadron colliders},''
  \href{http://dx.doi.org/10.1016/S0550-3213(98)00014-5}{{\em Nucl. Phys.} {\bf
  B515} (1998)  3--14},
\href{http://arxiv.org/abs/hep-ph/9710451}{{\tt arXiv:hep-ph/9710451}}.

\bibitem{LHC1}
M.~Lamont, ``{LHC near and medium terms prospects},''
  \href{http://arxiv.org/abs/{http://indico.desy.de/getFile.py/access?contribI%
d=8\&sessionId=15\&resId =1\&materialId=slides\&confId=1964}}{{\tt
  {http://indico.desy.de/getFile.py/access?contribId=8\&sessionId=15\&resId
  =1\&materialId=slides\&confId=1964}}}.

\bibitem{Amsler:2008zzb}
{\bf Particle Data Group} Collaboration, C.~Amsler {\em et al.}, ``{Review of
  particle physics},''
\href{http://dx.doi.org/10.1016/j.physletb.2008.07.018}{{\em Phys. Lett.} {\bf
  B667} (2008)  1}.

\bibitem{LHC2}
F.~Zimmermann, ``{LHC beyond 2020},''
  \href{http://arxiv.org/abs/{http://accnet.lal.in2p3.fr/Tasks/Literature/2010%
/KEK-Accelerator-Semin
  ar-14July2010-LHC-Beyond-2020-Frank-Zimmermann.pdf}}{{\tt
  {http://accnet.lal.in2p3.fr/Tasks/Literature/2010/KEK-Accelerator-Semin
  ar-14July2010-LHC-Beyond-2020-Frank-Zimmermann.pdf}}}.

\bibitem{Ellis:1983ed}
J.~R. Ellis and S.~Rudaz, ``{Search for Supersymmetry in Toponium Decays},''
\href{http://dx.doi.org/10.1016/0370-2693(83)90402-1}{{\em Phys. Lett.} {\bf
  B128} (1983)  248}.

\bibitem{Haber:1984rc}
H.~E. Haber and G.~L. Kane, ``{The Search for Supersymmetry: Probing Physics
  Beyond the Standard Model},''
\href{http://dx.doi.org/10.1016/0370-1573(85)90051-1}{{\em Phys. Rept.} {\bf
  117} (1985)  75--263}.

\bibitem{Bartl:1986hp}
A.~Bartl, H.~Fraas, and W.~Majerotto, ``{Production and Decay of Neutralinos in
  e+ e- Annihilation},''
\href{http://dx.doi.org/10.1016/0550-3213(86)90104-5}{{\em Nucl. Phys.} {\bf
  B278} (1986)  1}.

\bibitem{Rolbiecki:2009hk}
K.~Rolbiecki, J.~Tattersall, and G.~Moortgat-Pick, ``{Measuring the Stop Mixing
  Angle at the LHC},''
\href{http://arxiv.org/abs/0909.3196}{{\tt arXiv:0909.3196 [hep-ph]}}.

\bibitem{Choi:1999cc}
S.~Y. Choi, H.~S. Song, and W.~Y. Song, ``{CP phases in correlated production
  and decay of neutralinos in the minimal supersymmetric standard model},''
  \href{http://dx.doi.org/10.1103/PhysRevD.61.075004}{{\em Phys. Rev.} {\bf
  D61} (2000)  075004},
\href{http://arxiv.org/abs/hep-ph/9907474}{{\tt arXiv:hep-ph/9907474}}.

\bibitem{Fischer:2001gp}
M.~Fischer, S.~Groote, J.~G. Korner, and M.~C. Mauser, ``{Complete angular
  analysis of polarized top decay at O(alpha($s$) )},''
  \href{http://dx.doi.org/10.1103/PhysRevD.65.054036}{{\em Phys. Rev.} {\bf
  D65} (2002)  054036},
\href{http://arxiv.org/abs/hep-ph/0101322}{{\tt arXiv:hep-ph/0101322}}.

\end{thebibliography}\endgroup

\end{document}